\documentclass[aps,prb,twocolumn,secnumarabic,
superscriptaddress,footinbib,notitlepage,showpacs,floatfix]{revtex4-1} 
\usepackage[utf8]{inputenc}
\usepackage{longtable}
\usepackage{morefloats}
\usepackage[dvips]{graphicx,color}
\usepackage[dvipsnames]{xcolor}
\usepackage{epsfig,graphicx,amsfonts,amsbsy}
\usepackage{amsmath,amsfonts,amsthm,amssymb}
\usepackage{url}
\usepackage{verbatim}
\usepackage[colorlinks=true,allcolors=blue]{hyperref}
\usepackage{array}
\usepackage{multirow}
\usepackage[normalem]{ulem}
\usepackage{tabularx}
\usepackage{multirow}
\usepackage{braket} 
\usepackage{dsfont}
\usepackage{bm}
\usepackage{natbib}
\usepackage{multibib}
\usepackage{bbold} 

\usepackage{titlesec}

\graphicspath{../Figures/}

\begin{document}


\title{Exchange interaction of hole-spin qubits in double quantum dots \\
in highly anisotropic semiconductors}
\author{Bence Het\'enyi}
\affiliation{Department of Physics, University of Basel, Klingelbergstrasse 82, CH-4056 Basel, Switzerland}
\author{Christoph Kloeffel}
\affiliation{Department of Physics, University of Basel, Klingelbergstrasse 82, CH-4056 Basel, Switzerland}
\author{Daniel Loss}
\affiliation{Department of Physics, University of Basel, Klingelbergstrasse 82, CH-4056 Basel, Switzerland} 
\date{\today}

\begin{abstract}
We study the exchange interaction between two hole-spin qubits in a double quantum dot setup  in a silicon nanowire in the presence of magnetic and electric fields. Based on symmetry arguments we show that there exists an effective spin that is conserved even in highly anisotropic semiconductors, provided that the system has a twofold symmetry with respect to the direction of the applied magnetic field. This finding facilitates the definition of qubit basis states and simplifies the form of exchange interaction for two-qubit gates in coupled quantum dots.  If the magnetic field is applied along a generic direction, cubic anisotropy terms act as an effective spin-orbit interaction introducing novel exchange couplings even for an inversion symmetric setup. Considering the example of a silicon nanowire double dot, we present the relative strength of these anisotropic exchange interaction terms and calculate the fidelity of the $\sqrt{\text{SWAP}}$ gate. Furthermore, we show that the anisotropy-induced spin-orbit effects can be comparable to that of the direct Rashba spin-orbit interaction for experimentally feasible electric field strengths. 
\end{abstract}

\maketitle


\section{Introduction}

Over the last two decades localized spins in quantum dots (QDs) became a promising candidate for scalable quantum computing\cite{loss:pra98,kloeffel:annurev13}. Electron spins confined in semiconductor heterostructures benefit from the feasibility of coherent control via electric-dipole-induced spin resonance (EDSR)\cite{golovach:prb06,nowack:sci07,nadjperge:nat10,schroer:prl11} and exchange based two-qubit gates\cite{petta:sci05,brunner:prl11,veldhorst:nat15}. On the other hand, besides charge noise and phonon induced decoherence, electrons are also exposed to fluctuating nuclear spins\cite{khaetskii:prl02,coish:prb04,coish:pss09}. 

Holes confined in quantum dots\cite{bulaev:prl05,bulaev:prl07} have recently attracted much attention due to the possibility of fast single-qubit control by virtue of  a strong spin-orbit interaction (SOI)\cite{hao:nlt10,kloeffel:prb13,li:nlt15,maurand:ncomm16,watzinger:ncomm18,terrazos:arX18,gao:adm20}, and slow decoherence owing to the suppressed hyperfine interaction\cite{fischer:prb08,brunner:sci09,fischer:prl10,maier:prb13,watzinger:ncomm18}. Single-shot readout\cite{vukusic:nlt18}, exchange-coupled quantum dots\cite{greilich:nphot11,hardy:nanotech19} and two-qubit gates\cite{hendrickx:nat20} have recently been realized in systems, where the heavy-hole (HH) and light-hole (LH) states are well separated.

As opposed to planar QDs, eigenstates of holes being strongly confined in more than one directions have significant contributions from both the HH and LH states\cite{sercel:prb90,csontos:prb09}. These systems benefit from an even stronger Rashba type of SOI that relies on the HH-LH mixing and is not suppressed by the fundamental band gap\cite{kloeffel:prb11,kloeffel:prb18}. In agreement with recent experiments\cite{hao:nlt10,higginbotham:prl14,brauns:prb16,
voisin:nlt16,maurand:ncomm16,wang:sst17,crippa:prl18,froning:apl18,sun:nlt18,devries:nlt18}, Si and Ge/Si core/shell nanowires (NWs) are particularly promising platforms for such low-dimensional hole systems. Remarkably, these NWs and QDs therein can be formed with a complementary metal-oxide-semiconductor (CMOS) compatible fabrication process\cite{jiang:edl09,voisin:nlt16,maurand:ncomm16,crippa:prl18,kuhlmann:apl18}, which indicates an exceptional scalability. Furthermore, both Si and Ge are bulk inversion symmetric, leading to a suppressed piezoelectric interaction between holes and phonons, and can be isotopically enriched, allowing to reduce the number of nuclear spins to nearly zero
\cite{tyryshkin:natmat12,veldhorst:nnano14,
muhonen:nnano14,sigillito:prl15}.

Two-qubit gates between hole-spin qubits in NW QDs can be implemented in different ways. For example, the qubits can be coupled over long distances via floating metallic gates \cite{trifunovic:prx12} or via the cavity photons of transmission-line resonators \cite{kloeffel:prb13,nigg:prl17} by harnessing the strong, direct Rashba SOI (DRSOI) \cite{kloeffel:prb11,kloeffel:prb18}. Nearby qubits, on the other hand, can be coupled by electrically controlling the wave function overlap and thereby inducing an exchange interaction. However, this important possibility has not been explored yet since the HH-LH mixing renders the interaction multifaceted.

In this paper, we address the question how the HH-LH mixing affects the form of the exchange interaction in tunnel-coupled QDs. This question is relevant not only for two-qubit operations but also for, e.g., the implementation of singlet-triplet qubits \cite{benjamin:pra01,levy:prl02,taylor:natphys05} and spin-to-charge readout schemes \cite{loss:pra98,ono:sci02,petta:sci05} with holes. 
In the most general case the form
\begin{equation}
H_{(1,1)} = \frac 1 4 \boldsymbol \sigma^L\cdot \bar{\mathbf J}\,  \boldsymbol \sigma^R + \frac 1 2 \left(\boldsymbol \Delta^L \cdot \boldsymbol \sigma^L +\boldsymbol \Delta^R \cdot \boldsymbol \sigma^R\right)
\label{eq:H11}
\end{equation}
needs to be assumed for the interaction between the qubit basis-states $\ket 0$ and $\ket 1$ of the left ($L$) and right ($R$) QDs, where $\bar{\mathbf J}$ is the exchange-matrix 
and the coefficients in the single-qubit part $\boldsymbol \Delta^{L(R)}$ are related to the $g$-tensor $\bar{\mathbf g}^{L(R)}$ via $\boldsymbol \Delta^{L(R)} = \mu_B\, \bar{\mathbf g}^{L(R)} \mathbf B$. The Pauli-matrices (e.g., $\sigma_z^{L} = \ket{0_L}\bra{0_L}-\ket{1_L}\bra{1_L}$) are acting on the energetically lowest two eigenstates of the QDs. 

For electrons, the exchange matrix obtains the simple form $J_{ij} = J R_{ij}(\mathbf n, \theta)$, where the parameters of the rotation $R_{ij}$ depend on the spin-orbit couplings\cite{kavokin:prb04}. However due to the strong anisotropy of the hole states in materials like silicon, the exchange interaction acquires anisotropic corrections even in the case of an inversion symmetric setup. As reported earlier, these anisotropic effects can enhance the role of spatial symmetries in the theoretical description
\cite{wenk:prb16,venitucci:prb18,crippa:prl18}.

We discuss the possible symmetries of a generic double quantum dot (DQD) setup. If the confinement, the crystal structure, and the external fields respect the same symmetry, an effective spin can be associated to the qubit states of each QD. The conservation of this effective spin allows one to identify selection rules for the exchange interaction $\bar{\bf J}$.

We consider coupled hole-spin qubits in a silicon NW and identify the high symmetry axes of the magnetic field along which the effective spin is conserved. However, effective spin projections can get mixed upon application of an external electric field (inducing DRSOI), or by changing the direction of the magnetic field due to crystalline anisotropy (anisotropy-induced spin mixing). One of our central results is shown in Fig.~\ref{fig:exchange}, where we present the effect of the hitherto neglected anisotropy-induced spin mixing mechanism on the exchange interaction $\bar{\bf J}$ and the induced Zeeman splittings $\boldsymbol \Delta$. This mixing will lead to anisotropic corrections to the exchange interaction even in the presence of inversion symmetry. Furthermore, we compare our results obtained for silicon NWs with that of Ge/Si core/shell NWs, where crystalline anisotropy manifests itself rather weakly in the valence band.

If coupling is established between two QDs each hosting a single hole-spin qubit, the exchange interaction can be utilized to implement a fundamental entangling gate such as the $\sqrt{\text{SWAP}}$ \cite{loss:pra98}. However, in silicon the anisotropic corrections can lead to systematic gate errors limiting the fidelity of the $\sqrt{\text{SWAP}}$ gate. We calculate the gate fidelities in the coherent system and find that anisotropic corrections can be mitigated if the gate is sufficiently fast. 

This paper is organized as follows: In Sec.~\ref{sec:singlequbit} we review the simple model of conduction band electrons and present the Hamiltonian of the valence band holes together with the commonly used axial approximation. In Sec.~\ref{sec:symmcons} we introduce an effective spin and discuss its advantages for the application as spin qubits. Projecting the Hamiltonian of coupled quantum dots to the low-energy basis in Secs.~\ref{sec:lowenbasis} and \ref{sec:twoqubit}, we study the selection rules that apply for the exchange interaction if the quantum dots respect a mutual twofold symmetry. In Sec.~\ref{sec:NWexchange} we propose a symmetry-decomposition of the Hamiltonian that reveals the different effective spin mixing terms, compare the spin mixing effect of the cubic anisotropy and the DRSOI, present the relative energy scales of the anisotropic corrections to the exchange interaction, and calculate the anisotropy-limited fidelities of a $\sqrt{\text{SWAP}}$ gate. We conclude with a few remarks and a short summary in Secs.~\ref{sec:discussion} and \ref{sec:conclusion}. Technical details are deferred to Apps.~\ref{sm:nfold}-\ref{sm:SiGe}.

\section{Single hole-spin qubit}
\label{sec:singlequbit}

We consider a single hole confined by electric gates either in a NW or in a two-dimensional hole gas in a heterostructure. Provided that the energy scale associated with the temperature is much lower than the orbital splittings, the hole will occupy the lowest orbital state. If the confinement is significantly stronger along one or two axes, the lowest state will retain only two-fold degeneracy in the absence of magnetic field due to the different effective masses corresponding to the HH and the LH states\cite{kloeffel:prb11, kloeffel:prb18}. Splitting of these eigenstates by magnetic field establishes an effective two-level system $\ket{0},\, \ket{1}$ to be referred to as hole-spin qubit later on.

First we consider the general Hamiltonian of a single quasiparticle, an electron ($+$) or a hole ($-$) confined in a QD
\begin{equation}
H_\text{QD} = H_\text{b}(\mathbf k, \hat{\mathbf J}) +H_Z (\hat{\mathbf J}, \mathbf B) \pm e\mathbf E\cdot \mathbf r + V_\text{QD}(\mathbf r) +H^\text{c} ,
\label{eq:QD}
\end{equation}
where $H_\text{b}(\mathbf k, \hat{\mathbf J})$ is the bulk Hamiltonian of either the conduction band or the valence bands with the vector operator $\hbar\, \hat{\mathbf J}$ combining the atomic orbital angular momentum [$l=0\, (1)$ for the conduction band (valence bands)] and the spin. The crystal-momentum including the vector potential $\mathbf A$ is $\hbar \mathbf k = -i\hbar \nabla + e \mathbf A$, where $e$ is the positive elementary charge. The Zeeman term $H_Z(\hat{\mathbf J}, \mathbf B)$ contains the spherical and anisotropic corrections coupling the magnetic field $\mathbf B = \nabla \times \mathbf A$ to the angular momentum $\hbar\, \hat{\mathbf J}$. The electric field is taken into account via the term $ \pm e \mathbf E\cdot \mathbf r$ and the inversion symmetric confinement potential $V_\text{QD}(\mathbf r)$, where  $\mathbf r$ is the position operator of the quasiparticle. The last term $H^\text{c}$ contains further corrections such as the Rashba and the Dresselhaus spin-orbit interaction (which are higher order terms in the multi-band perturbation theory) as well as the strain and the interface effects.

In the case of the conduction band the $s$-wave property of the Bloch-functions (i.e., zero orbital angular momentum) implies that the angular momentum components are given by the three Pauli matrices i.e., $\hbar\, \hat J_i = \frac{\hbar}{2} \sigma_i$. Since the Pauli matrices together with the identity matrix form a complete basis, the effective Hamiltonian of the conduction band electrons in a homogeneous magnetic field can be written in the simple form
\begin{eqnarray}
\begin{split}
H_\text{cond}& (\mathbf k) + H_Z(\hat{\mathbf J}, \mathbf B) =\frac{\hbar^2 k^2}{2m^*}+
g^*\mu_\text B \mathbf B \cdot \hat{\mathbf J}\, ,
\end{split}
\label{eq:conde}
\end{eqnarray}
with the effective band mass $m^*$ and $g$-factor $g^*$. The special property of the Pauli matrices then imply that the Hamiltonian of Eq.~\eqref{eq:conde} has continuous axial symmetry (the Hamiltonian commutes with the spin projection $\hat J_B$ along the magnetic field).

In the presence of magnetic field the Luttinger-Kohn Hamiltonian $H_\text{LK} + H_Z$ describing the top of the HH-LH bands for cubic crystals can be written as
\begin{eqnarray}
\begin{split}
H_\text{LK}& (\mathbf k, \hat{\mathbf J}) + H_Z(\hat{\mathbf J}, \mathbf B) =\\
&\frac{\hbar^2}{2m}\left[ \left( \gamma_1 + \frac{5}{2} \overline \gamma \right) k^2 - 2\overline \gamma (\mathbf k \cdot \hat{\mathbf J})^2 \right] \\
&+ (2\kappa + \overline \gamma) \mu_\text B \mathbf B \cdot \hat{\mathbf J}  + \Delta \gamma\,  K(\mathbf k, \hat{\mathbf J}) + 2 q \mu_\text B \mathbf B \cdot \hat{\mathcal J}
\end{split}
\label{eq:LK}
\end{eqnarray}
where $m$ is the bare electron mass, $\gamma_1$ is the first Luttinger parameter, $\overline \gamma = (2\gamma_2 + 3\gamma_3)/5$ is the averaged Luttinger parameter, and $\Delta \gamma = \gamma_3 - \gamma_2$ is the prefactor of the terms with cubic symmetry\cite{lipari:prl70,footnote:cubcorr,footnote:LKtypo,luttinger:pr55,luttinger:pr56}. The spin-$3/2$ vector operator $ \hbar\, \hat{\mathbf J}$ is combining the atomic orbital angular momentum ($l=1$) and the spin. The Zeeman part $H_Z(\hat{\mathbf J}, \mathbf B) = 2\kappa \mu_\text B \mathbf B\cdot \hat{\mathbf J} + 2 q \mu_\text B \mathbf B \cdot \mathbf{ \hat{\mathcal J}}$ is composed of the isotropic and anisotropic terms\cite{footnote:aniszeeman} with coefficients $\kappa$ and $q$, respectively. 

The first two terms of the Hamiltonian in Eq.~\eqref{eq:LK} are invariant under arbitrary rotations around the magnetic field axis i.e., $[H^{(1,2)},e^{-i \phi \hat F_B}] = 0$ holds for any angle $\phi$, where $\hbar\, \hat F_B$ is the total angular momentum $\hbar\hat{\bf F} =\hbar \hat{\bf J} +\hbar\hat{\bf L}$ projected along the magnetic field, with the orbital angular momentum being $\hbar\hat{\bf L} = -i\hbar\, \bf r \times \nabla$\cite{luttinger:pr56}. For materials like Ge, InAs, and GaAs these terms give the main contributions, since $\Delta \gamma \ll \overline \gamma $ and $q\ll |\kappa|$ (e.g., the anisotropy parameters $\Delta \gamma/\overline \gamma$ obtained from Ref.~[\onlinecite{winkler:book}] are $0.28,\, 0.091,$ and $0.31$, respectively), and the last two terms are treated only perturbatively within the framework of the \emph{axial approximation}\cite{sercel:prb90,csontos:prb09,doty:prb10,winkler:book,miserev:prb17}.

While for electrons even the spin $\hbar\, \hat J_B$ is approximately conserved, only the total angular momentum conservation could be considered for the valence band states. However, corrections due to cubic anisotropy can play important role\cite{terrazos:arX18}, especially for materials with strong cubic anisotropy (e.g., silicon where $\Delta \gamma / \overline \gamma = 1.1$) corrections to the axial approximation cannot be treated perturbatively. We wish to identify an effective spin as a good quantum number that is conserved by the Hamiltonian in Eq.~\eqref{eq:LK}, for the highly anisotropic case. For this  we consider the point symmetry group of the QD system in the next section.

\section{Symmetry considerations}
\label{sec:symmcons}

To properly define a qubit, we first consider the symmetries of the bulk crystal in the presence of magnetic and electric fields and identify high symmetry axes. This will allow us to identify an effective spin $\alpha = \text{mod}_2(F_B) \in \{-1/2, 1/2\}$\cite{footnote:modulo}, which is related to the eigenvalues of a twofold symmetry operator such as $D(C_{2B}) = e^{-i \pi \hat F_B}$, where $\hat{F}_B$ is the total angular momentum operator and has half-odd-integer eigenvalues $F_B$. The eigenstates in a QD are also characterized by this quantum number $\alpha$ and can be used as a qubit, provided that the confinement respects the considered symmetry. 
Finally, we present DQD geometries where a twofold symmetry is maintained implying spin selection rules for the interaction between the two quantum dots.  

The Bravais lattice of a bulk crystal is defined by discrete translations in the three spatial directions. The Bravais lattice can be invariant under further symmetry transformations, e.g., $N$-fold rotations $C_{Na}$ about an axis $a$, inversion $I$, or their combinations, the so called rotoreflections $S_{Na} = I\cdot C_{Na}$. The set of symmetry elements taking the lattice into itself constitute the point group of the crystal\cite{solyom:vol1}. The external fields can also be described in the language of point groups as follows. The homogeneous electric and magnetic fields $\bf E$   and $\bf B$ are invariant under any rotations around their axis. In addition, $\bf E$ is symmetric and under reflections with respect to any mirror plane that contains its axis. However, since $\bf B$ is a pseudo-vector it only respects inversion symmetry and reflection symmetry with respect to the single mirror plane being perpendicular to it.

Comparing the symmetries of a cubic crystal with that of the external fields one obtains the reduced point group of the crystal in the presence of external fields which we summarize in Tab.~\ref{tab: sym} for different directions of the external fields. The resulting point group is non-trivial, only if the magnetic field is applied along a high-symmetry axis, e.g., the point group $C_4$, which contains the elements of a four-fold rotation around the axis of the magnetic field $\mathbf B$, i.e., $C_4 = \{E,C_{4B},C_{2B},C^3_{4B}\}$, where $E$ is the identity element.

\begin{table}[h]
\centering
\begin{tabular}{|c|c|c|c|c|}
 \hline 
& $B \parallel \braket{100}$ & $B \parallel \braket{110}$ & $B \parallel \braket{111}$& other \\ 
 \hline 
 $E = 0$ & $C_{4h}\rightarrow  \alpha_4$ & $C_{2h}\rightarrow \alpha$ & $C_{3i}\rightarrow  \alpha_3$ & $C_i$ \\ 
 \hline 
 $E\parallel B$ & $C_4\rightarrow \alpha_4$ & $C_2\rightarrow \alpha$ & $C_3\rightarrow  \alpha_3$ & $C_1$  \\ 
 \hline 
 $E\perp B$ & $C_s\rightarrow \alpha$ & $C_s\rightarrow \alpha$ & $C_1$ & $C_1$  \\ 
 \hline 
 \end{tabular} 
\caption{Reduction of the cubic point group (using Schoenflies symbols) with diamond structure $O_h$ upon application of external electric and magnetic fields\cite{winkler:book}. To each point group containing an $N$-fold symmetry one can associate (indicated by an arrow) a generalized effective spin $\alpha_{N} = \text{mod}_N (F_B)$ as discussed in App.~\ref{sm:nfold}. The special case of a two-level system in the ground state of a single QD with $\alpha\equiv\alpha_{2}$ is used here as qubit basis.}
\label{tab: sym}
\end{table}

\begin{figure}
\centering
\includegraphics[width=\columnwidth]{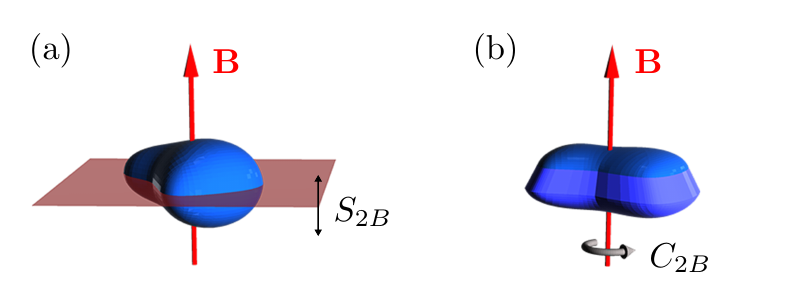}
\caption{Sketch of single QDs with two-fold symmetry $\mathcal R_2$ in the presence of a magnetic field $\bf B$ (red) where the blue objects illustrate the shape of the QDs in real space, e.g., the geometry of the confinement or the charge density of the confined holes.
(a) The QD possesses a mirror symmetry $\mathcal R_2 = S_{2B}$; the only direction of the magnetic field respecting the symmetry of the QD is perpendicular to the symmetry plane (red transparent). 
(b) The QD possesses a twofold rotation symmetry $\mathcal R_2 = C_{2B}$; the only direction of the magnetic field respecting the symmetry is along the symmetry axis (red). }
\label{fig:symmetriessQD}
\end{figure}

We have seen, that the bulk crystal can have a non-trivial point group even if external fields are applied. Moreover, if we consider a quasiparticle confined in a QD, the point group of this system consists of the symmetry elements that respect the symmetries of the crystal, the fields, and the confinement (i.e., the intersection of the corresponding point groups). Fig.~\ref{fig:symmetriessQD} illustrates confinement geometries respecting only a single twofold symmetry of the magnetic field. The resulting point group is $S_2 = \{E,S_{2B}\}$ for the system in Fig.~\ref{fig:symmetriessQD}(a) and $C_2 = \{E,C_{2B}\}$ for Fig.~\ref{fig:symmetriessQD}(b).

If $\mathcal R_{2} \in \{C_{2B},S_{2B}\}$ is a twofold symmetry element of the point group of the QD system, the Hamiltonian of Eq.~\eqref{eq:QD} has to commute with the symmetry operator $D(\mathcal R_2)$, the representation of  $\mathcal R_{2}$ on the Hilbert space\cite{solyom:vol1}. As a consequence, the (non-degenerate) eigenstates $\ket m \in \{\ket 0 , \ket 1, \ket 2,...\}$ of the Hamiltonian are also eigenstates of $D(\mathcal R_2)$,
\begin{equation}
D(\mathcal R_{2}) \ket{m\alpha} = e^{-i \pi \alpha} \ket{m\alpha}\, ,
\label{eq:effspin}
\end{equation}
where $\alpha = \text{mod}_2 (F_B)$ is a spin-like quantum number of the state $\ket{m\alpha} \equiv \ket m$, where, again, $\alpha = \pm 1/2$. Furthermore, it can be shown that the two states of a Kramers doublet (states that are transformed to each other by time-reversal) correspond to effective spin $\alpha$ and $-\alpha$ (see App.~\ref{sm:gs}). With this finding we conclude that spin qubits can be defined as the lowest Kramers doublet of a quantum dot in any crystal or confinement geometry, as long as a twofold symmetry is preserved in the system.

The effective spin $\alpha$  is rooted in the discrete rotational (rotoreflectional) symmetry and gives rise to a discrete conservation law for the total angular momentum expressed as $\alpha =  \text{mod}_2(F_B)$. This relation can be proven in general for an $N$-fold symmetry axis (with  $N\geq 2$) which gives rise to a quantum number $\alpha_N =  \text{mod}_N(F_B)$ that is conserved modulo $N$. This result can be seen as Bloch's theorem for angular momenta (see App.~\ref{sm:nfold} for further details).

The interaction between holes with their effective spin gives rise to matrix elements of the interaction which satisfy certain selection rules due to the underlying conservation laws. In particular,  we find that the matrix elements for the Hamiltonian of coupled QDs $H_\text{DQD}$ obey the following selection rules:
\begin{equation}
 \braket{m \alpha, n \beta |H_\text{DQD}| p \chi, q \xi} \propto \delta_{0,\text{mod}_2(\alpha+\beta - \chi -\xi)}\, ,
\label{eq:selection}
\end{equation}
provided that the DQD setup respects the twofold symmetry of the left and right QDs (for details see App.~\ref{sm:VCzeros}). The two-particle states above, 
$\ket{ m \alpha, n \beta}= \ket{m \alpha}_1\otimes \ket{ n \beta}_2$, are product states of the single-particle states $\ket{m\alpha}_1$ and $\ket{n\beta}_2$ for the first and second particle, respectively. The indices $m,n,p,q \in \{0_L,0_R,1_L,1_R,2_L,...\}$ label the single-particle eigenstates of the left or right QD, and $ \alpha, \beta, \chi, \xi =  \pm 1/2$ stand for the effective spins associated with the single-particle states. 

\begin{figure}
\centering
\includegraphics[width=1\columnwidth]{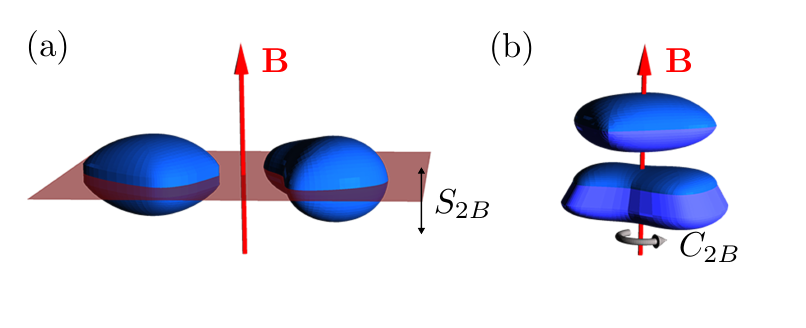}
\caption{DQD geometries where the symmetry-induced quantum numbers $\alpha_{L(R)}=\pm 1/2$ are conserved and can be used to label a qubit in each QD. The DQD axis (a) lies in the symmetry plane (red transparent plane) that  is perpendicular to the magnetic field $\bf B$ or (b) coincides with both the magnetic field and the twofold rotation axis.}
\label{fig:symmetriesDQD}
\end{figure}

We have seen that the point group of the crystal may contain two-, three- or fourfold symmetry axes, even if external fields are applied. However, the confinement potential defining the DQD should also respect these symmetries in order to benefit from the selection rules  given in Eq.~\eqref{eq:selection}. A DQD setup can obey a twofold symmetry in two ways: {(i)} The DQD axis (i.e., the axis connecting the centers of the two coupled QDs) lies in the common symmetry plane of the QDs being perpendicular to the magnetic field [as illustrated in Fig.~\ref{fig:symmetriesDQD}(a)]. {(ii)} The DQD axis coincides with the common rotation axis, see Fig.~\ref{fig:symmetriesDQD}(b).

\section{Low-energy basis of a DQD}
\label{sec:lowenbasis}

In order to determine the interaction between the two qubits in a DQD system the low-energy solutions of the following Hamiltonian\cite{footnote:partexchsym} have to be considered,
\begin{equation}
H_\text{DQD} = H_L(1)+\delta V_L(1)+ H_R(2)+\delta V_R(2) + C(1,2)\, ,
\label{eq:DQD}
\end{equation}
where $H_{L(R)}$ is the single-QD Hamiltonian in Eq.~\eqref{eq:QD} of the left (right) QD with $V_\text{QD} = V_{L(R)}$, $\delta V_{L(R)} = V_\text{DQD} - V_{L(R)}$ is the difference between the double-  and single-dot potentials, and $C(1,2)=e^2/(4\pi \epsilon |\mathbf r_1-\mathbf r_2|)$ is the Coulomb interaction with the single-particle coordinates $\mathbf r_{1,2}$ and the dielectric constant $\epsilon=\epsilon_0 \epsilon_r$, where $\epsilon_0$ is the vacuum permittivity.

In order to construct a basis for the low-energy effective Hamiltonian of the DQD, the energetically lowest eigenstates (for qubits) $\ket{0_L},\, \ket{1_L}$ of the single QD Hamiltonian $H_L$ need to be orthonormalized with respect to the ones in the right well $\ket{0_R},\, \ket{1_R}$\cite{burkard:prb99}. In general, these eigenstates can not be written as a product of an orbital and a spin part\cite{burkard:prl02,stepanenko:prb03} (e.g., due to spin-orbit interaction or the HH-LH mixing) and therefore the $\ket 0$ and $\ket 1$ eigenstates of different QDs are not necessarily orthogonal. To characterize this mutual non-orthogonalities, we introduce the  overlap matrix elements
\begin{equation}
S_{ab}  = \braket{a_L| b_R}\, ,
\label{eq:Sab}
\end{equation}
where $a,b \in \{0,1\}$. Even though the wave functions of electrons in the presence of inversion symmetry is separable into orbital and spin part, ensuring $S_{0 1} =0$ and $S_{0 0} =S_{1 1}$ regardless of the magnetic field, for holes these relations hold only for zero magnetic field\cite{kavokin:prb04}. Nevertheless we find that $S_{0 1} $ can also vanish provided the magnetic field preserves a twofold symmetry in the DQD system, while the difference of the diagonal elements is proportional to the applied magnetic field, i.e., $S_{0 0} - S_{1 1} \propto B$ for small magnetic fields.

The overlap $S_{ab}$ is suppressed exponentially with distance between the dots, therefore the orthonormalization can be performed in such a way that the same quantum numbers can be used to label the orthonormalized states, e.g., $\ket{0_L}_\text{ON} = \sum_{m} C_{m,0_L} \ket{m}$, where $C_{m,0_L} \sim \delta_{m,0_L} + \mathcal{O}(S_{ab})$ (the precise form is given in App.~\ref{sm:orth_methods}). Further on the subscript ``ON" will be suppressed for simplicity, and the non-orthogonal states are used only in the definition in Eq.~\eqref{eq:Sab}.

From the orthonormalized low-energy single-particle states, six fermionic two-particle states can be constructed in accordance with the Pauli principle. Three of them are analogous to triplet states, $\ket{T_{0\pm}}$ [e.g., $\ket{T_+} = (\ket{0_L, 0_R} -\ket{0_R, 0_L} ) / \sqrt 2 $], one corresponds to a singlet state with a single particle on each dot $\ket S$, and there are two singlet states $\ket{S_{L(R)}}$ with both particles on the left (right) dot. Within the framework of the Hund-Mulliken approximation these six states are used to project the Hamiltonian of Eq.~\eqref{eq:DQD} onto the low-energy Hilbert space.

\section{Low-energy Hamiltonian of a DQD}
\label{sec:twoqubit}

Exchange coupling is known to be adequate for implementation of two-qubit gates\cite{loss:pra98}. However the effect of the doubly occupied singlets ($\ket{S_{L(R)}}$) can be crucial for correct quantitative analysis\cite{burkard:prb99}, for simplicity we restrict the DQD Hamiltonian of Eq.~\eqref{eq:DQD} to the lowest energy subspace $\{S,T_0,T_+,T_- \}$ for the qualitative discussion here and take the higher singlets into account only for the numerical results in Sec.~\ref{sec:NWexchange} via Schrieffer-Wolff transformation\cite{bravyi:annp11}.

Changing the confinement of the left QD (initially described by the Hamiltonian $\frac{1}{2} \Delta^L_0\, \sigma^L_z$) results in a modification of the Zeeman splittings $\boldsymbol \Delta^L$ inducing coupling between the qubit basis states\cite{venitucci:prb18,crippa:prl18}. To lowest order in the potential difference $\delta V_L = V_\text{DQD}-V_L$ one obtains 
\begin{subequations}
\begin{equation}
\Delta^L_z =\Delta^L_0 + \braket{0_L|\delta V_L|0_L} - \braket{1_L|\delta V_L|1_L} \, ,
\label{eq:oz}
\end{equation}
\begin{equation}
\Delta^L_x-i \Delta^L_y = 2\braket{0_L|\delta V_L|1_L}\, ,
\label{eq:oxy}
\end{equation}
\end{subequations}
for the coefficients of the single-qubit effective Hamiltonian $\frac{1}{2} \boldsymbol \Delta^L \cdot \boldsymbol \sigma^L$, where the Pauli matrices $\sigma^L_{x,y,z}$ are defined with the orthonormalized states $\ket{a_L}$ (e.g., $\sigma^L_{z} = \ket{0_L}\bra{0_L}- \ket{1_L}\bra{1_L}$). If the DQD respects the same twofold symmetry as the left and right QDs, the couplings vanish, i.e., $\Delta^L_x = \Delta^L_y =0$ due to the modulo-2 conservation law. On the other hand, the energy splitting $\Delta^L_z$ can still be affected by the potential. The results given by Eqs.~\eqref{eq:oz}~and~\eqref{eq:oxy} are in correspondence with those obtained from the study of mirror symmetries in the $g$-matrix formalism\cite{venitucci:prb18}. 

Next, we turn to a discussion of the exchange couplings. First we point out that, in the presence of a twofold symmetry, due to Eq.~\eqref{eq:selection} the $T_\pm$-triplet sector is decoupled from the rest of the subspace. This is so because the triplets $\ket{T_\pm}$  are composed from  products of two single-particle states with the same quantum number $\alpha$, whereas all the other states contain products of single-particle states with opposite quantum numbers $\alpha$. While this decoupling also exists for conduction band states, the matrix elements $\braket{T_+|C|T_-}$ and $\braket{S|C|T_0}$ (which vanish for electrons) do not need to vanish for valence band holes.

In the most general case, the Hamiltonian of two coupled qubits is given in Eq.~\eqref{eq:H11}. Rewriting this expression in the singlet-triplet basis $\{S,T_0,T_+,T_- \}$, we obtain the effective Hamiltonian
\begin{widetext}
\begin{eqnarray}
\begin{split}
H_{(1,1)} = \frac{1}{4} \begin{pmatrix}
-J_{xx}-J_{yy}-J_{zz} & 2iJ^a_{xy} & \sqrt 2 \left(-J^a_{xz} - i J^a_{yz}\right) & \sqrt 2 \left(-J^a_{xz} + i J^a_{yz}\right)  \\
-2i J^a_{xy} & J_{xx}+J_{yy}-J_{zz} & \sqrt 2 (J^s_{xz} + i J^s_{yz}) & -\sqrt 2 (J^s_{xz} - i J^s_{yz}) \\
\sqrt 2 \left(-J^a_{xz} + i J^a_{yz}\right) &  \sqrt 2 (J^s_{xz} - i J^s_{yz}) & J_{zz} & J_{xx}-J_{yy} -2iJ^s_{xy} \\
\sqrt 2 \left(-J^a_{xz} - i J^a_{yz}\right) & \sqrt 2 (-J^s_{xz} - i J^s_{yz}) & J_{xx}-J_{yy} +2iJ^s_{xy} & J_{zz}\\
\end{pmatrix}\\
+\frac 1 2 \begin{pmatrix}
0 & 2 \Delta^a_z & \sqrt 2 \left(-\Delta^a_x - i \Delta^a_y\right) & \sqrt 2 \left(\Delta^a_x - i \Delta^a_y\right)  \\
2 \Delta^a_z & 0 & \sqrt 2 \left(\Delta^s_x + i \Delta^s_y\right) & \sqrt 2 \left(\Delta^s_x - i \Delta^s_y\right) \\
\sqrt 2 \left(-\Delta^a_x + i \Delta^a_y\right) &  \sqrt 2 \left(\Delta^s_x - i \Delta^s_y\right) &  2\Delta^s_z & 0 \\
\sqrt 2 \left(\Delta^a_x + i \Delta^a_y\right) & \sqrt 2 \left(\Delta^s_x + i \Delta^s_y\right) & 0 & - 2 \Delta^s_z \\
\end{pmatrix}\, ,
\end{split}
\label{eq:H11long}
\end{eqnarray}
\end{widetext}
where $\bar{\mathbf J}^s=\big(\bar{\mathbf J}+\bar{\mathbf J}^T\big)/2$ is the symmetric part, $\bar{\mathbf J}^a=\big(\bar{\mathbf J}-\bar{\mathbf J}^T \big)/2$ is the antisymmetric part of the exchange matrix, and the Zeeman terms $\boldsymbol  \Delta^s =(\boldsymbol \Delta^L + \boldsymbol \Delta^R)/2$ account for the homogeneous part of the magnetic field, while $ \boldsymbol \Delta^a = (\boldsymbol \Delta^L - \boldsymbol \Delta^R)/2$ for the inhomogeneous part.
In general, these terms can also arise if the g-factors of the dots are different. If the magnetic field is oriented along a high-symmetry axis, the $T_\pm$ sector becomes independent of the $ST_0$ sector, and therefore the off-diagonal elements of the exchange matrix $J^s_{xz},J^a_{xz},J^s_{yz},$ and $J^a_{yz}$ have to vanish.

\section{Coupled hole-spin qubits in silicon NWs}
\label{sec:NWexchange}

In this section we consider a cylindrical NW fabricated from silicon. A coupled QD setup is established by means of electrostatic gates such that each QD is occupied by a single hole. First we discuss what assumptions were made and which parameter values were used to describe the system, then in Sec.~\ref{subsec:effspincons} we decompose the Hamiltonian of a single QD into an effective spin conserving and a symmetry breaking part. We compare the effect of different mechanisms that can lead to anisotropic exchange interaction in Sec.~\ref{subsec:Efield}, namely the DRSOI and the anisotropy-induced spin mixing (to be clarified below). In Sec.~\ref{subsec:anis} we provide numerical examples for the parameters characterizing the effective $4\times 4$ Hamiltonian in Eq.~\eqref{eq:H11long} for the inversion symmetric limit, where the anisotropic exchange couplings can only appear by virtue of the strong cubic anisotropy in silicon. Finally, the effect of anisotropic corrections on the fidelity of a $\sqrt{\text{SWAP}}$ gate is discussed in Sec.~\ref{subsec:SWAP}.

To be concrete for the numerical evaluations to follow, we focus on a silicon NW  with circular cross section and cylinder axis along the $[001]$ direction of the silicon crystal. We use $\gamma_1 = 4.285$, $\gamma_2 = 0.339$, $\gamma_3 = 1.446$, $\kappa = -0.42$, and $q=0.01$ for the Luttinger-parameters and $\epsilon_r = 12.1$ for the dielectric constant\cite{winkler:book}. The axes $x,y,z$ correspond to the $[100]$, $[0 1 0]$, $[001]$ crystallographic axes, respectively. The magnetic field $\mathbf B = B(\cos \varphi,\sin \varphi,0)$ is applied in the plane perpendicular to the cylinder axis, and it is parametrized by the angle $\varphi$ it encloses with the $x$ axis [see Fig.~\ref{fig:overlaps}(a)].

For the confinement potential in Hamiltonian of Eq.~\eqref{eq:DQD} we assume the form $V_\text{DQD}(\mathbf r) = V_\text{NW} (x,y)+\frac{v_B}{a^4} (z^2-a^2)^2$ with $v_B$ being the height of the quartic-potential barrier, and $V_\text{NW} (x,y)$ is the transverse  confinement potential. In order to efficiently approximate the wells of the DQD potential by independent harmonic potentials $V_{L(R)}(z)$, the barrier height should be larger than the orbital energy of the harmonic confinement, or equivalently $v_B> 2\, \hbar^2 \gamma_1/(m a^2)$ ($ \sim 2.9\, \text{meV}$ for the parameters of our example of Si NW). The confinement potential $V_\text{DQD}(\mathbf r)$ is inversion symmetric [due to the cylindrical shape of the NW, i.e. $V_\text{NW}(x,y) = V_\text{NW}(x^2+y^2)$] and the inversion-asymmetric part of the electric fields is taken into account via the term $-e\mathbf E\cdot \mathbf r$, where $\mathbf E$ is a homogeneous electric field.

\subsection{Corrections beyond the conservation of effective spin}
\label{subsec:effspincons}

Since the magnetic field $\mathbf{B}$ in the present case is always perpendicular to the DQD axis ($z$ axis), only reflection symmetry $S_{2B}$ can be maintained [see Fig.~\ref{fig:symmetriesDQD}(a)]. For example, if the magnetic field is applied along the $[100]$ axis ($\varphi = 0$) and $\mathbf E = 0$, the system respects the symmetry $S_{2B}$, facilitating the definition of the effective spin $\alpha$. The modulo-2 conservation of this effective spin simplifies the form of the exchange interaction matrix and the induced Zeeman splittings (as we discussed in Sec.~\ref{sec:twoqubit}).

Changing the direction of the magnetic field or the application of a homogeneous electric field can break the symmetry. Since the confinement potential $V_\text{DQD} (\mathbf r)$ and the Coulomb interaction $C(\mathbf r_1-\mathbf r_2)$ both respect the symmetry $S_{2B}$ for any $\varphi$, the symmetry breaking contribution in the Hamiltonian of Eq.~\eqref{eq:DQD} has to be a part of the single QD Hamiltonian $H_\text{QD}$. In this case, the mixing of effective spins has a strong similarity with the case of localized electron spins in the presence of  Rashba or Dresselhaus spin-orbit interaction\cite{burkard:prl02,stepanenko:prb03}.  Therefore, we are motivated to decompose the single QD Hamiltonian in the following way,
\begin{subequations}
\label{eq:decomp}
\begin{equation}
H_0 = H_\text{QD} - H_\text{SO}\, ,
\end{equation}
\begin{equation}
H_\text{SO} = \frac 1 2 [D(S_{2B}),H_\text{QD}]D(S_{2B})\, ,
\end{equation}
\end{subequations}
where $H_0$ commutes with the symmetry operator $D(S_{2B})$ conserving the effective spin, and the analogue of the SOI, $H_\text{SO}$, anti-commutes with the symmetry and thus leads to couplings only of  sectors with different quantum numbers. 

Performing the decomposition given in Eq.~\eqref{eq:decomp} on the Hamiltonian of Eq.~\eqref{eq:QD}, one obtains the following three terms for the spin-non-conserving part $H_\text{SO}$:
\begin{subequations}
\begin{equation}
\begin{split}
\frac{\hbar^2}{2m} \Delta \gamma \sin{(4\varphi)}&\left[ (k_\perp^2-k_B^2)\{\hat J_B,\hat J_\perp\} \right.\\
&\left. + \{k_B,k_\perp\}(\hat J_\perp^2-\hat J_B^2) \right]\, ,
\end{split}
\label{eq:anissoia}
\end{equation}
\begin{equation}
\begin{split}
2q\mu_\text{B} B \frac{\sin{(4\varphi)}}{4} \left[\hat J_\perp(\hat J_\perp^2-\hat J_B^2)- 2 \hat J_B \{\hat J_B,\hat J_\perp\} \right]\, ,
\end{split}
\label{eq:anissoib}
\end{equation}
\begin{equation}
-e E_B r_B\, ,
\label{eq:anissoic}
\end{equation}
\end{subequations}
where $\{A,B\}=(AB+BA)/2$ defines the anti-commutator. The momenta rotated to the frame of the magnetic field are defined as $k_B = k_x \cos{(\varphi)} + k_y \sin{(\varphi)}$ and $k_\perp = -k_x \sin{(\varphi)} + k_y \cos{(\varphi)}$, analogously $\hbar\, \hat J_B$ and $\hbar\, \hat J_\perp$ are the rotated angular momenta and $r_B$ is the rotated coordinate. The first term, Eq.~\eqref{eq:anissoia} is coming from the momentum-resolved part of the LK Hamiltonian, the second, Eq.~\eqref{eq:anissoib} is related to the anisotropic Zeeman term, and the third term, Eq.~\eqref{eq:anissoic} arises from the electric field component $E_B$ perpendicular to the symmetry plane, i.e., parallel to the magnetic field $\mathbf B$.

Note that the anisotropic spin-orbit corrections of Eqs.~\eqref{eq:anissoia}~and~\eqref{eq:anissoib} are proportional to $\sin{(4\varphi)}$ and therefore vanish if $\mathbf B \parallel [100],\,\, \mathbf B \parallel [110],$ etc., in correspondence with the expectations from the symmetry arguments. Below we will see that this oscillatory $\varphi$-dependence manifests itself in the overlap matrix element $S_{01}$ [see Figs.~\ref{fig:overlaps}(b)-(c)], the induced Zeeman splittings $\Delta_{x,y}$, and the off-diagonal exchange couplings $J_{xy},J_{xz}$ and $J_{yz}$ [see Figs.~\ref{fig:exchange}(b)-(c)].

\subsection{Effects of homogeneous electric fields on the effective spin mixing} 
\label{subsec:Efield}

\begin{figure*}
\centering
\includegraphics[width=0.37\textwidth]{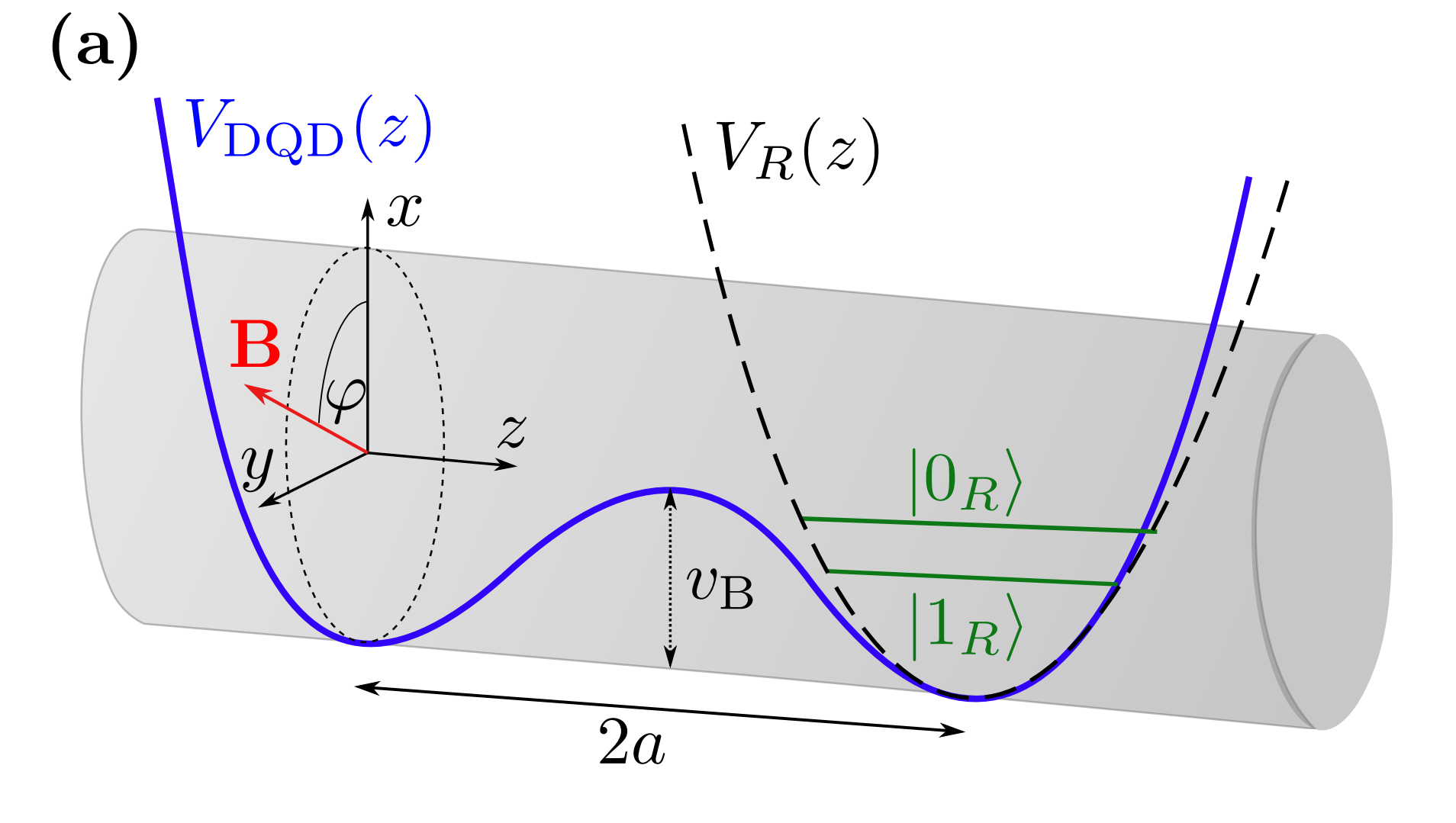}
\includegraphics[width=0.62\textwidth]{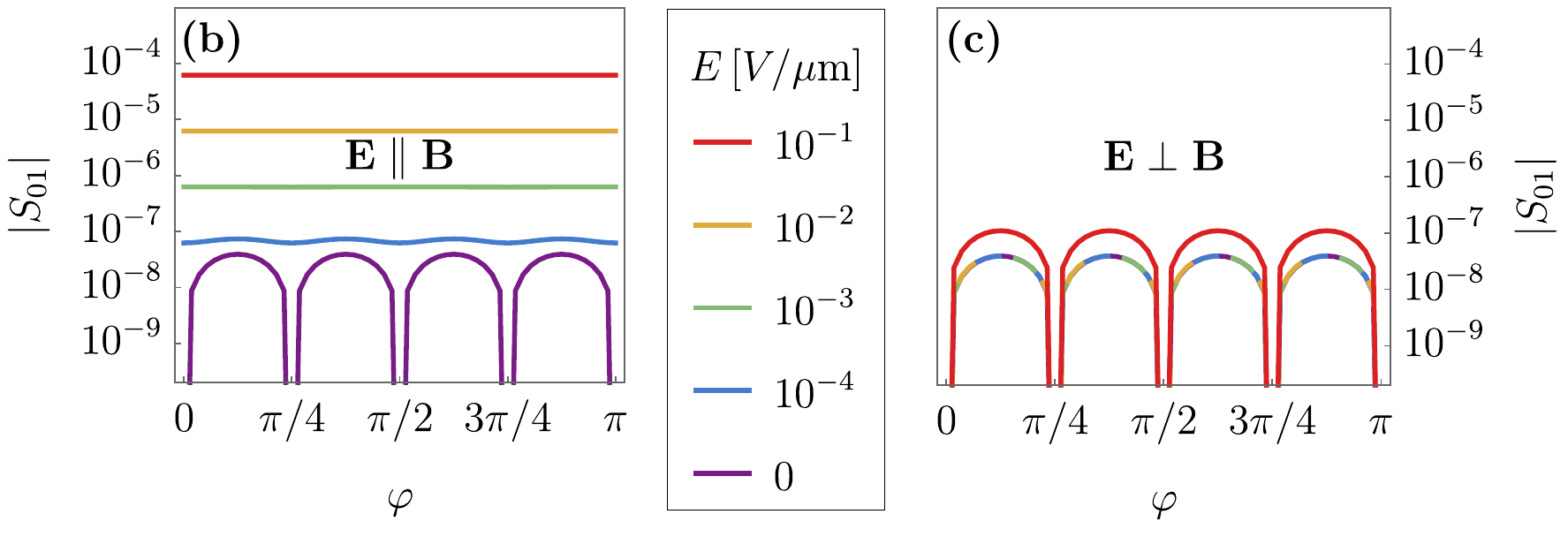}
\caption{(a) Schematic figure of the DQD system realized in a silicon NW. (b)-(c) Absolute value of the anti-aligned overlap $|S_{01}| = |\braket{0_L|1_R}|$, as a function of $\varphi$, the angle enclosed by the magnetic field $\mathbf B = B(\cos \varphi,\sin \varphi,0)$ and the $x \parallel [100]$ axis, (b) when the electric field $\mathbf E = E(\cos \varphi,\sin \varphi,0)$ is applied parallel to the magnetic field, and (c) when applied perpendicular to the magnetic field, $\mathbf E = E(-\sin \varphi,\cos \varphi,0)$. For this calculation we used a distance of $2a = 30\, \text{nm}$ between the QDs, barrier height of $v_\text B = 3\, \text{meV}$, magnetic field of $B = 1\, \text T$, and cylindrical hard-wall confinement with a radius of $R = 7\, \text{nm}$. Details on the basis choice and adopted assumptions can be found in App.~\ref{sm:basis}.}
\label{fig:overlaps}
\end{figure*}

As we have seen for the spin-non-conserving part $H_\text{SO}$ of the Hamiltonian $H_\text{QD}$, the symmetry $S_{2B}$ can be broken  by terms with cubic anisotropy in Eqs.~\eqref{eq:anissoia}~and~\eqref{eq:anissoib} or due to a finite electric field component $E_B$ along the magnetic field in Eq.~\eqref{eq:anissoic}. As a consequence, the qubit states
of different QDs are no longer orthogonal to each other. In order to qualify and compare these two anisotropy effects, we take the overlap matrix element $|S_{01}|=|\braket{0_L | 1_R }|$ as a figure of merit for this qubit mixing, since it is usually nonzero but vanishes when the QDs respect the symmetry $S_{2B}$. Furthermore, $S_{01}$ can be shown to be proportional to the off-diagonal exchange interaction $J_{xz}$ and $J_{yz}$ and the induced Zeeman splittings $\Delta_x$ and $\Delta_y$ (see App.~\ref{sm:S01}). 

For the results presented in Figs.~\ref{fig:overlaps}(b)-(c) we used a cylindrically symmetric hard-wall confinement in the transverse  directions for $V_\text{NW}(x,y)$  and studied the effect of asymmetries via the homogeneous electric field term $-e\mathbf E  \cdot \mathbf r$ with $\mathbf E = (E_x,E_y,0)$. The standard Rashba SOI is also taken into account with the coefficient $\alpha_\text{h} = 0.002\, \text{nm}^2 e$ (according to Ref.~[\onlinecite{kloeffel:prb18}]), although the effect of this term is dominated by the DRSOI\cite{kloeffel:prb11,kloeffel:prb18}. For the example above the relative deviation from the $\alpha_\text{h} = 0$ case is less than $1\%$.

In Fig.~\ref{fig:overlaps}(b) a homogeneous electric field $\mathbf E$ is applied parallel to the magnetic field, i.e., $\mathbf E \cdot \mathbf r = E_B r_B$. We plot $|S_{01}|$ as a function of the magnetic field direction $\varphi$ for different strength of the electric field between $0$ and $0.1\,V/\mu \text m$. Even for relatively small electric fields, e.g., $E_B \sim 10^{-3}\, V/\mu \text m$, the overlap $|S_{01}|$ becomes independent of $\varphi$ and changes roughly linearly with the electric field $E_B$. These findings are in agreement with the strong DRSOI predicted for this growth direction\cite{kloeffel:prb11,kloeffel:prb18}.

In Fig.~\ref{fig:overlaps}(c) the electric field is applied perpendicular to the magnetic field and the wire axis, therefore $E_B = 0$ and the symmetry $S_{2B}$ can only be broken by the cubic anisotropy terms. Importantly, $S_{01}$ obtains the same angular dependence $\sin{(4\varphi)}$ as the terms of the spin non-conserving part of the Hamiltonian $H_\text{SO}$ (see also  App.~\ref{sm:S01}). Unlike the DRSOI contribution, this effect does not have an analogue in the case of conduction band electrons, since it appears even in the presence of inversion symmetry ($\mathbf E = 0$). We refer to this phenomenon as  anisotropy-induced spin mixing (see below).

\subsection{Anisotropic exchange interaction in the presence of inversion symmetry} 
\label{subsec:anis}

\begin{figure*}
\centering
\includegraphics[width=\textwidth]{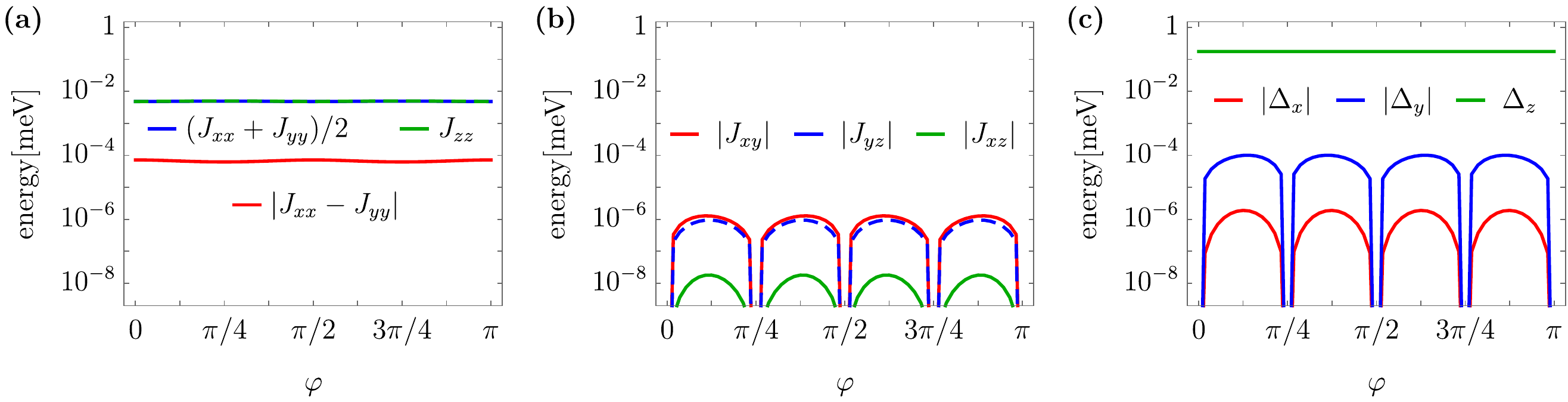}
\caption{(a)-(c) Coefficients characterizing the exchange matrix $J_{ij}$ and the single-particle Hamiltonian $\Delta_i$ as a function of the magnetic field direction $\varphi$, for a silicon NW. For the numerical simulation the following parameters were used: DQD distance $2a = 30\, \text{nm}$; barrier height $v_\text B = 3\, \text{meV}$; magnetic field $B = 1\, \text T$; harmonic potential of Eq.~\eqref{eq:Vharm} with a confinement length of $2l_T = 8\, \text{nm}$. For details on the basis states used for the numerics see App.~\ref{sm:basis}.}
\label{fig:exchange}
\end{figure*}

In the absence of the homogeneous electric field, i.e., $\mathbf E = 0$, the Hamiltonian of Eq.~\eqref{eq:DQD} is inversion-symmetric, implying that the low-energy Hamiltonian of Eq.~\eqref{eq:H11} is invariant under the swap of $L$ and $R$. The exchange matrix has to be symmetric ($\bar{\mathbf J} = \bar{\mathbf J}^s $) and the Zeeman splittings have to be identical ($\boldsymbol \Delta^a = 0$), therefore in line with Eq.~\eqref{eq:H11long} the inversion symmetry decouples the singlet $\ket S$ from the three triplets, but  the $\ket{T_0}$ state can still be coupled to the $\ket{T_\pm}$ states in the Hamiltonian $H_{(1,1)}$. 

To simplify the numerical calculation of the Coulomb integrals, the eigenstates of the left (right) QD $\ket{0_{L(R)}}$ and $\ket{1_{L(R)}}$ were calculated numerically using harmonic confinement,
\begin{equation}
V_\text{NW} (x,y) = \frac{\hbar^2 \gamma_1}{2m l^4_T} (x^2+y^2),
\label{eq:Vharm}
\end{equation}
for the transverse directions as well, where $2l_T$ is the diameter of the NW. After the orthonormalization, the two-particle states were constructed in order to project the Hamiltonian of Eq.~\eqref{eq:DQD} to the lowest $6\times 6$ subspace. In order to take the effect of the doubly occupied singlets into account, we perform a second-order Schrieffer-Wolff transformation and obtain the coefficients of the effective Hamiltonian in Eq.~\eqref{eq:H11long} as a function of the magnetic field direction $\varphi$. The result\cite{footnote:gauge} is presented in Fig.~\ref{fig:exchange}.

As implied by Eq.~\eqref{eq:selection}, the off-diagonal elements of the exchange matrix $J_{xz}$ and $J_{yz}$ and the Zeeman splittings (i.e., $\Delta_{x,y}$) corresponding to off-diagonal terms in the single-qubit Hamiltonian vanish, if $\mathbf B$ is along a high-symmetry direction. The coupling between the $\ket{T_\pm}$ states, i.e., $\braket{T_-|C|T_+} =\frac 1 4 (J_{xx} - J_{yy} + 2iJ_{xy})$ remains finite regardless of the angle $\varphi$. The exchange matrix element $J_{xy}$ vanishes along the high-symmetry directions in Fig.~\ref{fig:exchange}, however, this is only due to the relative phase between the numerically calculated basis states $\ket{T_\pm}$. The $ST_0$ splitting $(J_{xx}+J_{yy})/2$ is approximately equal to $J_{zz}$, but this feature is observed only for small enough potential barriers $v_B$. The anisotropic exchange matrix elements and the Zeeman splittings acquire their highest value at the low-symmetry field direction $\varphi \sim \pi/8$,  i.e., in-between high-symmetry points.

Having obtained the effective interaction between the two qubits, we are now in the position to  discuss how  the anisotropic corrections affect the fidelity of  two-qubit gates. This will be done in the following subsection.

\subsection{$\sqrt{\text{SWAP}}$ gate with anisotropic exchange interaction}
\label{subsec:SWAP}

\begin{figure*}
\centering
\includegraphics[width=\textwidth]{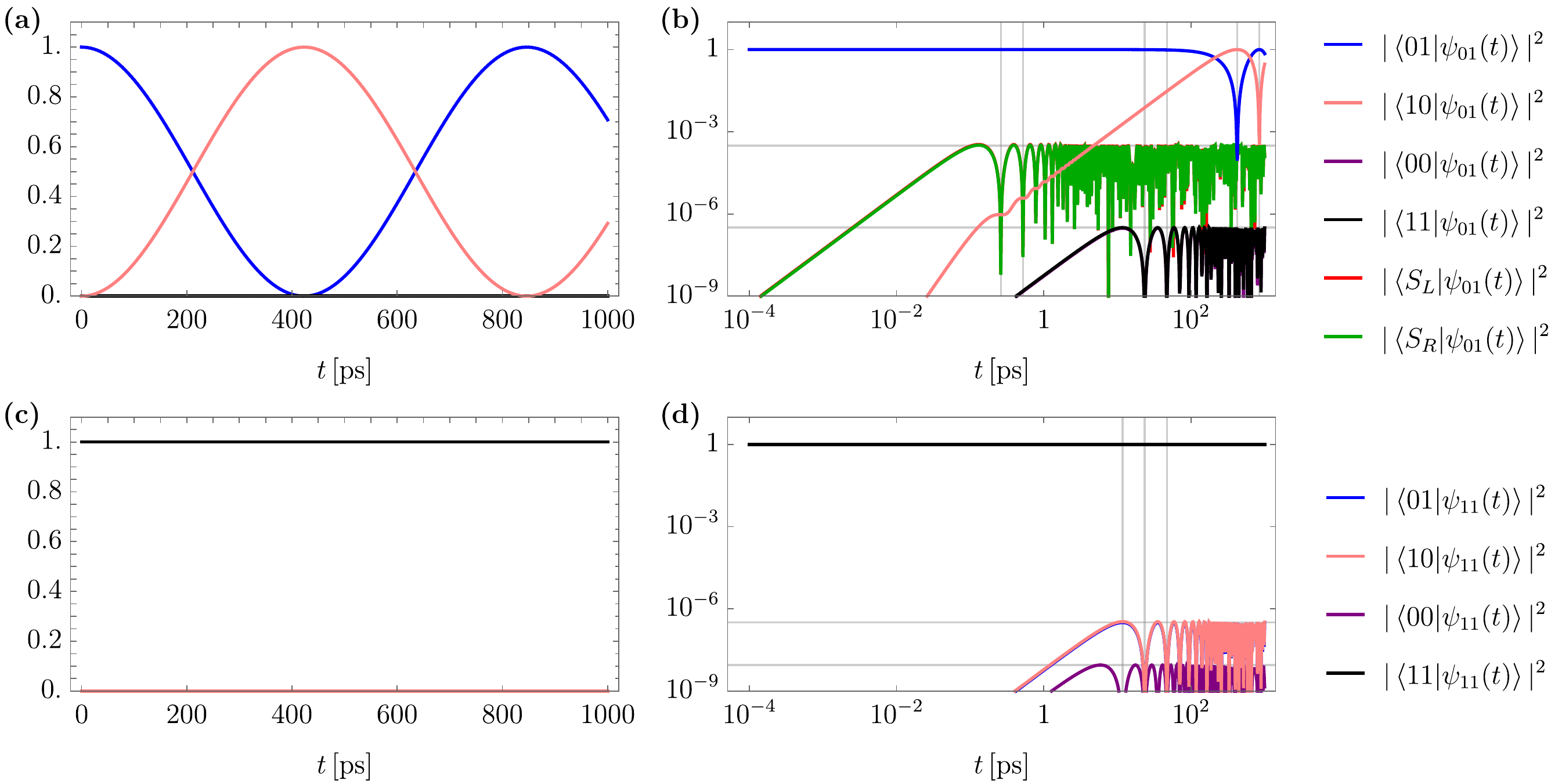}
\caption{Overlaps between DQD basis states and the states $\ket{\psi_{01}(t)}$  [(a) and (b)] and $\ket{\psi_{11}(t)}$  [(c) and (d)] as a function of time for the low-symmetry case where $\varphi = \pi/8$, for a silicon NW. The time evolution is shown (a) on a liner-linear scale (b) on a log-log scale, with the horizontal lines showing the estimates for the overlaps obtained in Eqs.~\eqref{eq:SWAP1}-\eqref{eq:SWAP3} [(c) and (d) similarly]. Horizontal lines are showing the maximal overlap as a function of time [note that the one corresponding to $|\braket{01|\psi_{11}(t)}|^2 = |\braket{11|\psi_{01}(t)}|^2$ is shown on both (b) and (d)]. Overlaps with different basis states are oscillating with the half-cycle duration of $\sim h/U$ for $|\braket{S_L|\psi_{01}(t)}|^2$,  $\sim h/\Delta_z$ for $|\braket{11|\psi_{01}(t)}|^2$, and  $\sim h/(2\Delta_z)$ for $|\braket{00|\psi_{11}(t)}|^2$ as illustrated by the vertical lines in (b) and (d).}
\label{fig:SWAPerrors}
\end{figure*}

Isotropic exchange interaction is a well-known way to implement the $\sqrt{\text{SWAP}}$ gate\cite{loss:pra98}  from which the fundamental CNOT gate can be obtained. However, in Fig.~\ref{fig:exchange} we have seen that anisotropic exchange matrix elements and off-diagonal Zeeman splittings (i.e., $\Delta_{x,y}$) emerge which might affect the operation of such a quantum gate for exchange coupled hole-spin qubits.

In this subsection we adopt the notation widely used in the literature of quantum computation. Instead of the two-particle states with $(1,1)$ charge configuration we introduce the two-qubit basis states $\ket{00} = \ket{T_+}$, $\ket{11} = \ket{T_-}$, $\ket{01} = (\ket{T_0} - \ket S)/\sqrt 2 $, and $\ket{10} = (\ket{T_0} + \ket S)/\sqrt 2 $. However, the discussion of the gate errors for exchange coupled QDs cannot be complete without taking into account the doubly occupied singlets $\ket{S_L}$ and $\ket{S_R}$ explicitly. An ideal $\sqrt{\text{SWAP}}$ gate leaves the qubit states unchanged, if the two qubits are in the $\ket{00}$ or $\ket{11}$ state, while creating a maximally entangled state, if they are either in the $\ket{01}$ or in the $\ket{10}$ state, e.g., $U_{\sqrt{\text{\tiny SWAP}}} \ket{01} = \ket{-} \equiv (\ket{01} - i \ket{10})/\sqrt 2$. 

Since the exchange interaction is electrically tunable via the potential barrier $v_B$\cite{loss:pra98,burkard:prb99,burkard:prb00,
reed:prl16,martins:prl16}, we consider a case where initially the two qubits are in a (disentangled) product state e.g., $\ket{01}$ or $\ket{11}$ and the interaction is switched on at $t=0$. Therefore, we study the time evolution of the state $\ket{\psi_{01} (t)}$ according to the effective low-energy $6\times 6$ Hamiltonian $H_\text{DQD}^{6\times 6}$ corresponding to the parameters of Sec.~\ref{subsec:anis}, such that the time evolution starts from a product state, e.g., $\ket{\psi_{01} (0)} = \ket{01}$. The state $\ket{\psi_{01} (t)}$ will have the highest overlap with the target state $\ket{-}$, when $t = \tau_s = \hbar \pi\, (J_{xx} + J_{yy})^{-1}$. To benchmark the accuracy of the gate corresponding to the $\ket{01}$ input state, we define the fidelity $\mathcal F_{01} = |\braket{-|U_{\sqrt{\text{\tiny SWAP}}} |01}|^2 = |\braket{-|\psi_{01} (\tau_s)}|^2$ and the error rate $1-\mathcal F_{01}$.

To illustrate the role of the anisotropic corrections, we compare the performance of the gate for two different magnetic field directions, a high symmetry case ($\varphi = 0$), where the system respects the twofold symmetry $S_{2B}$, and a low-symmetry case ($\varphi = \pi/8$), where the anisotropic corrections are the largest in Fig.~\ref{fig:exchange}. Due to the relatively low potential barrier $v_B = 3\, \text{meV}$, fast operation times $\tau_s\sim 210\,\text{ps}$ can be achieved, but the fidelity $\mathcal F_{01}$ is limited by the tunneling to the doubly occupied states [see Figs.~\ref{fig:SWAPerrors}(a)-(b) for the low-symmetry case]. Exploiting the $L\leftrightarrow R$ symmetry, the error rate $1-\mathcal F_{01}$ can be simply estimated by
\begin{equation}
1-\mathcal F_{01} \sim 2 |\braket{S_L|\psi_{01}(\tau_s)}|^2 \sim \frac{J_{xx}+J_{yy}}{U}\, ,
\label{eq:SWAP1}
\end{equation}
where $U \sim 15\,\text{meV}$ is the charging energy. This error also sets a limit for the fidelity in the high-symmetry case, since the singlet-singlet tunneling cannot be ruled out by symmetry arguments.

Due to the anisotropic coupling terms in Hamiltonian $H_\text{DQD}^{6\times 6}$, the $\ket{11}$ and $\ket{00}$ states are also affected by the operation. Introducing the state $\ket{\psi_{11} (t)}$, such that the time evolution starts from a product state, i.e., $\ket{\psi_{11} (0)} = \ket{11}$, we define the fidelity of the gate corresponding to the $\ket{11}$ input state as $\mathcal F_{11} = |\braket{11|U_{\sqrt{\text{\tiny SWAP}}} |11}|^2 = |\braket{11|\psi_{11} (\tau_s)}|^2$ and the corresponding error rate $1-\mathcal F_{11}$. In Figs.~\ref{fig:SWAPerrors}(c)-(d) the time evolution of the overlaps of the state $\ket{\psi_{11}(t)}$ with the four two-qubit basis states are shown in the low-symmetry case ($\varphi = \pi/8$). In this case, the fidelity is limited by the transition probability from the input state $\ket{11}$ to the $\ket{01}$ and $\ket{10}$ states. The estimated error rate is then given by the induced Zeeman splittings as
\begin{equation}
1-\mathcal F_{11} \sim 2 |\braket{01|\psi_{11}(\tau_s)}|^2 \sim 2 \frac{\Delta_{x}^2+\Delta_{y}^2}{\Delta_z^2}\, .
\label{eq:SWAP2}
\end{equation}
As pointed out in Sec.~\ref{sec:twoqubit}, the $\ket{00}$ and $\ket{11}$ states ($\ket{T_\pm}$ states in the earlier notation) are coupled to each other even in the presence of a twofold symmetry. The error rate $1-\mathcal F_{11}$ in the high-symmetry case is then determined by the anisotropic correction to the exchange term $J_{xx}-J_{yy}$ as follows
\begin{equation}
1-\mathcal F_{11} \sim |\braket{00|\psi_{11}(\tau_s)}|^2 \sim \left( \frac{J_{xx}-J_{yy}}{4\Delta_z}\right)^2\, .
\label{eq:SWAP3}
\end{equation}
However, for the system considered in Sec.~\ref{subsec:anis} the fidelity $\mathcal F_{11}$ is significantly higher than $\mathcal F_{01}$ for both the high- and low-symmetry cases (see Tab.~\ref{tab:SWAP} for the calculated values), implying that for low enough potential barriers $v_B$ the fidelity of the $\sqrt{\text{SWAP}}$ gate is not limited by the anisotropic corrections but by the probability of tunneling to a doubly occupied state (the opposite limit with a high potential barrier $v_B$ is discussed in App.~\ref{sm:timevo}).

\begin{table}
\begin{tabular}{|c|c|c|}
\hline 
&&\\[-1em]
 & high symmetry  & low symmetry  \\ 
\hline 
&&\\[-1em]
$\varphi$ & $0$ & $\pi /8$ \\ 
&&\\[-1em]
\hline 
&&\\[-0.9em]
$1-\mathcal F_{01}$ & $6.2\cdot 10^{-4}$ & $6\cdot 10^{-4}$ \\ 
\hline 
&&\\[-0.9em]
$1-\mathcal F_{11}$ & $ 10^{-8}$ & $6.4\cdot 10^{-7}$ \\ 
\hline 
\end{tabular}
\caption{Error rates of the $\sqrt{\text{SWAP}}$ gate for the high-symmetry case ($\varphi = 0$) and a low-symmetry case ($\varphi=\pi/8$, providing the poorest fidelities as a function of magnetic field direction) for the silicon NW setup illustrated in Fig.~\ref{fig:overlaps}(a).}
\label{tab:SWAP}
\end{table}

\section{Discussion}
\label{sec:discussion}

{\it Validity of the axial limit.} In the axial limit ($\Delta \gamma,\, q \rightarrow 0$), the symmetry of the Hamiltonian in Eq.~\eqref{eq:LK} is higher than the actual symmetry of the system. In this case due to the continuous rotation symmetry of the Hamiltonian, the total angular momentum component $\hbar F_B$ $(\, = \hbar \alpha_\infty)$ is a good quantum number regardless of the magnetic field direction. Therefore this approximation completely ignores the couplings between $T_\pm$ and the $ST_0$ sector that arise even for perfectly inversion symmetric confinement. 

Since the anisotropic terms coupling states with different quantum numbers in Eqs.~\eqref{eq:anissoia}~and~\eqref{eq:anissoib} are proportional to $\Delta \gamma$ and $q$, they are expected to be suppressed for materials of lower anisotropy. A comparison of exchange interaction between the above studied silicon and the Ge/Si core/shell NW presented in App.~\ref{sm:SiGe} is consistent with this expectation.

{\it Silicon NW with different arrangements.} In Ref.~[\onlinecite{kloeffel:prb18}] the authors suggested to study silicon NWs with $\braket{100}$ growth direction, since a Rashba-type of SOI is enhanced in these directions compared to the conventionally used $\braket{110}$ growth direction\cite{footnote:growthdirection}. On the other hand, the $\braket{110}$ growth direction can be advantageous for the study of the anisotropy induced SOI effects, since it is less susceptible to  external electric fields.

Another interesting feature of the $\braket{100}$ growth direction is that the NW axis coincides with a 4-fold symmetry axis. If the magnetic field is applied along the NW, the modulo-$4$ conservation law of the angular momentum rules out the anisotropic coupling $J_{xx}-J_{yy}$ as well. However, in order to achieve such a symmetry in an experimental setup, the electrostatic gates would have to be arranged such that they respect the 4-fold rotation symmetry and therefore this favourable case does not seem to be within current experimental reach.

{\it Orientation of the spin-orbit vector.} When an electron or hole propagates along the NW axis $z$, an electric field along the $x$ direction induces an effective magnetic field along the $y$ axis on account of Rashba SOI. For holes, however, additional terms can arise such that the electric-field-induced effective magnetic field (spin-orbit vector) is not parallel to $y$. For example, considering a silicon NW with $z \parallel \braket{100}$ and a square cross-section, the calculations in Ref.~[\onlinecite{kloeffel:prb18}] resulted in an effective magnetic field whose component along $x$ (parallel to the electric field) is nonzero unless $\gamma_2 = \gamma_3$ or $\sin (4\phi') = 0$, where the angle $\phi'$ depends on the orientation of the crystallographic axes with respect to the NW cross-section. We note that this result has remarkable similarities with Eq.~\eqref{eq:anissoia}. The symmetry considerations in the present work provide a simple and intuitive explanation for the unusual, effective magnetic field component parallel to the electric field derived in Ref.~[\onlinecite{kloeffel:prb18}].

{\it Further signatures for the $T_\pm$ decoupling.} As discussed in Sec.~\ref{sec:twoqubit}, a remarkable consequence of the conserved quantum number is the vanishing exchange interaction and single-particle couplings between the $T_\pm$ sector and the remaining four basis states. Besides the numerical results for the overlap $S_{01}$ and the matrix elements in Fig.~\ref{fig:exchange}, we also studied the crossing of the $\ket{S_L}$ and the $\ket{T_-}$ energy levels as a function of detuning and the leakage current near the crossing point. These results also confirmed the decoupling of the sectors with different quantum numbers.

{\it Magnetic Weyl points.} In Refs.~[\onlinecite{scherubl:cp19}-\onlinecite{frank:arX19}], the authors find that topologically protected magnetic degeneracy points (referred to as magnetic Weyl points) can appear in DQDs for arbitrary SOI, i.e., for certain orientations of the magnetic field $\pm \mathbf B_W$, the singlet $\ket{S}$ and the lower triplet state $\ket{T_-}$ become degenerate: the levels cross as function of magnetic field and are protected from  hybridization.

In the DQD system considered here, the decoupling of the $T_\pm$ states from the $ST_0$ states also leads to magnetic degeneracy points at fine-tuned magnetic fields. These degeneracies are protected by the two-fold (in general $N$-fold) symmetry of the DQD system. 
However, establishing a connection between the topologically and symmetry protected magnetic degeneracies requires further analysis.

{\it Applications in experiments.} Our results corroborate the strong anisotropy in spin-related quantities observed in recent experiments\cite{wang:nlt16,voisin:nlt16,brauns:prb16,hung:prb17}. Furthermore, many of the recent experimental setups seem to be invariant under reflection with respect to a certain plane, e.g., the plane being perpendicular to the plunger gates in Refs.~[\onlinecite{froning:apl18}]~and~[\onlinecite{scherubl:cp19}]. Since the orientation of the magnetic field is usually tunable via a two- or three-dimensional vector magnet, the study of anisotropic effects is well within the reach of state-of-the-art experiments. {\it In situ} control of the confinement is usually also available by all-electrical means via tuning the confinement gate voltages\cite{camenzind:2019}.

\section{Conclusion}
\label{sec:conclusion}

We showed that an effective spin quantum number can be assigned to confined hole-states even in highly anisotropic materials if the magnetic field is applied along a twofold symmetry axis of the system.  Even though in general, the isotropic Heisenberg exchange is not sufficient to describe the interaction, exchange based two-qubit gates are likely to be feasible for hole-spin qubits in NWs. Besides enabling fast single-qubit operations by purely electrical means\cite{kloeffel:prb13}, silicon and Ge/Si core/shell NWs are also promising platforms to realize fast and high fidelity two-qubit operations.

\acknowledgments

We thank A. P\'alyi, M. Russ, and S. Bosco for helpful discussions. This work was supported by  the Swiss National Science Foundation and NCCR QSIT. 


\appendix

\section{Magnetic field along an $N$-fold symmetry axis, generalization of Bloch's theorem}
\label{sm:nfold} 

Considering a system with continuous translational symmetry and imposing periodic boundary conditions with a period of $L$, the eigenstates of the Hamiltonian are characterized by the momentum $p = \hbar\, n\frac{2\pi}{L}$ with $n \in \mathbb Z$.  For discrete translational symmetry Bloch's theorem\cite{bloch:zp29} states that only the wave number $k =\text{mod}_G (p/ \hbar)$ is a good quantum number, where $G = \frac{2\pi} a$, is the primitive reciprocal lattice vector with $a$ being the lattice constant. Even though the wave number $k$ is not a consequence of a continuous symmetry, it obeys the conservation law $\text{mod}_G[\sum_i (k_i -k'_i)]=0$ for microscopic processes involving more than one particle, where $k_i$ and $k'_i$ are the initial and final wave numbers of the $i$th particle.

An analogous conservation law holds for the angular momentum in a system with discrete rotational symmetry. The proof is very similar to that of Bloch's theorem for a one-dimensional lattice, only the translation operator $T_a = \exp{(-i\hat p a/\hbar)}$, with $\hat p$ being the momentum operator, needs to be replaced by the rotation operator $R_\phi = \exp{(-i\phi \hat F_z)}$, with $\hbar \hat F_z$ being the $z$-component of the total angular momentum operator, $\hbar \hat {\bf F} = \hbar \hat {\bf L} + \hbar \hat {\bf J}$, keeping in mind that for particles with half-odd-integer spin the periodic boundary condition should be imposed for $\phi = 4\pi$. 

For a system where $z$ is an $N$-fold symmetry axis the eigenstates of the Hamiltonian are characterized by the quantum number $\alpha_N = \text{mod}_N(F_z)$, where $F_z$ is the associated angular momentum eigenvalue.
A derivation analogous to the case of translational symmetry leads to the conservation of the quantum number $\alpha_N$, namely 
\begin{equation}
\text{mod}_N \Big[\sum_i (\alpha_{Ni}-\alpha_{Ni}') \Big] = 0\, ,
\end{equation}
where $\alpha_{Ni}$ and $\alpha_{Ni}'$ are the initial and final quantum numbers of the $i$th particle. The wave function $f_{m,j,\alpha_N} ({\bf r}) = \braket{{\bf r},j | m\alpha_N}$ corresponding to the single particle eigenstate $\ket{m\alpha_N}$ of the Hamiltonian respecting an $N$-fold symmetry can be written as
\begin{equation}
 f_{m,j,\alpha_N}(r,\phi,z)
=e^{i (\alpha_N-j) \phi}
\,u_{m,j,\alpha_N} (r,\phi,z)\, ,
\label{smeq:rotwf}
\end{equation}
where $j$ is the eigenvalue of  $ \hat J_z$, and the function $u_{m,j,\alpha_N} ({\bf r})$ respects the $N$-fold rotation symmetry, i.e., $u_{m,j,\alpha_N}(r,\phi,z) = u_{m,j,\alpha_N}(r,\phi+2\pi n/N,z)$ for every $n \in \mathbb Z$. In the $N\rightarrow \infty$ limit $\alpha_\infty = F_z$ is conserved in agreement with Noether's theorem and Eq.~\eqref{smeq:rotwf} corresponds to the ansatz of the axial limit from Refs.~[\onlinecite{sercel:prb90}-\onlinecite{kloeffel:prb11}]. This finding also explains the presence or absence of ``hole-spin mixing" for vertically stacked lateral QDs in Ref.~[\onlinecite{doty:prb10}], even without assuming cylindrical symmetry for the LK Hamiltonian. 

In the case of rotoreflections $S_{NB}$, the derivation is very similar, however, the symmetry of the function $u_{m,j,\alpha_N} ({\bf r})$ corresponds to the rotoreflection, i.e., $u_{m,j,\alpha_N}(r,\phi,z) = u_{m,j,\alpha_N}(r,\phi+\pi n+2\pi n/N,(-1)^n z)$ for every $n \in \mathbb Z$.

\section{Lowest-energy Kramers doublet}
\label{sm:gs}

In this appendix we show that the two lowest-energy eigenstates of a single-particle Hamiltonian $H_{B=0}$ obeying a twofold symmetry $\mathcal R_2$ have to belong to different values of the quantum number $\alpha\equiv \alpha_2$ in the absence of a magnetic field.

According to Kramers theorem, in the absence of magnetic field each energy level should be at least twofold degenerate. Let us assume that there are no fields and the two energetically lowest eigenstates  $\ket{\psi_1}$ and $\ket{ \psi_2} = \mathcal T \ket{\psi_1}$ are degenerate ground states, i.e.,
\begin{equation}
H_{B=0} \ket{\psi_{1,2}} = \varepsilon_0 \ket{\psi_{1,2}}\, ,
\end{equation}
with $\mathcal T$ being the anti-unitary time-reversal operator and $\varepsilon_0$ is the ground state energy. Furthermore, due to the anti-unitarity of $\mathcal T$ the time-reversed partner $\ket{\psi_2}$ is necessarily orthogonal to $\ket{\psi_1}$\cite{solyom:vol1}.

If the Hamiltonian commutes with a twofold symmetry operator $D(\mathcal R_2)$, the eigenstates $\ket{\psi_{1,2}}$ can always be chosen to be simultaneous eigenstate of $D(\mathcal R_2)$ as well,
\begin{equation}
D(\mathcal R_2)\ket{\psi_1} = \pm i \ket{\psi_1}\, .
\end{equation}
with the eigenvalue either $+i$ or $-i$. Since $\mathcal T$ commutes  with  $D(\mathcal R_2)$, one obtains
\begin{equation}
\begin{split}
D(\mathcal R_2) \ket{\psi_2} &=D(\mathcal R_2)\mathcal T\ket{\psi_1} =\mathcal T D(\mathcal R_2)\ket{\psi_1}\\ 
&=\mathcal T \left(\pm i \ket{\psi_1} \right) =\mp i \mathcal T \ket{\psi_1}=\mp i \ket{\psi_2}\, .
\end{split}
\end{equation}
In other words, if $\ket{\psi_1}$ belongs to the quantum number $\alpha = +1/2$ its time reversed partner $\ket{\psi_2}$ has to have the opposite quantum number $\alpha = -1/2$. 

Consequently, when the doublet is split by an external magnetic field such that the twofold symmetry is preserved, the resulting eigenstates are of different quantum number (unless the Zeeman splitting exceeds the orbital splitting, in which case the lowest eigenstates of $H_\text{QD}$ are not time-reversed partners of each other).

\section{Conservation of the effective spin}
\label{sm:VCzeros}

In this appendix we derive Eq.~\eqref{eq:selection} of the main text. First we list some important relations regarding the effect of the symmetry operator $D_2 \equiv D(\mathcal R_2)$ on single- and two-particle states which are eigenstates of this operator:
\begin{subequations}
\begin{equation}
D_2 \ket{m\alpha} = e^{-i\pi \alpha} \ket{m\alpha},
\end{equation}
\begin{equation}
D_2^2 \ket{m\alpha} = - \ket{m\alpha},
\end{equation}
\begin{eqnarray}
\begin{split}
\tilde D_2 \ket{m\alpha,n \beta} &= \left( D_2^{(1)}\otimes D^{(2)}_2 \right) \ket{m\alpha, n\beta}\\
& = (-1)^{\alpha + \beta} \ket{m\alpha, n\beta},
\end{split}
\end{eqnarray}
\begin{equation}
\tilde D_2^2 \ket{m\alpha, n\beta} = \ket{m\alpha, n\beta},
\end{equation}
\end{subequations}
where the two-particle states above, 
$\ket{ m \alpha, n \beta}= \ket{m \alpha}_1\otimes \ket{ n \beta}_2$, are product states of the single-particle states $\ket{m\alpha}_1$ and $\ket{n\beta}_2$ for the first and second particle, respectively. The indices $m,n,p,q \in \{0_L,0_R,1_L,1_R,2_L...\}$ label the single-particle eigenstates of the left or right QD, and $ \alpha, \beta, \chi, \xi =  \pm 1/2$ stand for the effective spins associated to the single-particle states. The representation of the symmetry $\tilde D_2$ acts on the two-particle states as a tensor product of the corresponding single-particle representations $D_2^{(1,2)}$. Furthermore, we see that $\tilde D_2^2$ acts on the two-particle states as the identity, in correspondence with the fact that these states are always of integer spin.

Provided that the DQD Hamiltonian $H_\text{DQD}$ commutes with the symmetry operator $\tilde D_2$ using the relations given above one finds
\begin{eqnarray}
\begin{split}
\bra{m\alpha, n\beta}H_\text{DQD}&\ket{p\chi, q\xi} = \braket{m\alpha, n\beta|H_\text{DQD} \tilde D_2^2|p\chi, q\xi}\\
&= \braket{m\alpha, n\beta|\tilde D_2\, H_\text{DQD}\, \tilde D_2|p\chi, q\xi}\\
&= (-1)^{\alpha+\beta +\chi+\xi} \braket{m\alpha, n\beta|H_\text{DQD}|p\chi, q\xi}\, .
\end{split}
\end{eqnarray}
Subtracting the rightmost part of the equation from the leftmost, we obtain
\begin{equation}
(1-(-1)^{\alpha+\beta +\chi+\xi}) \braket{m\alpha, n\beta|H_\text{DQD}|p\chi, q\xi} =0\, ,
\end{equation}
which leads to Eq.~\eqref{eq:selection} in the main text.

\section{Orthonormalization method}
\label{sm:orth_methods}

In this appendix we present a method to orthonormalize the single-particle states of the left $\ket{a_L}$ and right dot $\ket{b_R}$. Even though the Gram-Schmidt procedure is a well known method for the orthonormalization, being a recursive method it can not ensure the $L\leftrightarrow R$ symmetry for the orthonormalized states. Here we present a method that is although less straightforward, conserves the physically relevant implications of the inversion symmetry. 

Assuming $L\leftrightarrow R$ symmetry, the most general transformation connecting the single-particle states $\ket{a_{L(R)}}$ to the orthonormalized states $\ket{a_{L(R)}}_\text{ON}$ has a block-matrix structure, i.e.,
\begin{equation}
\begin{pmatrix}
\ket{0_L}_\text{ON}\\
\ket{1_L}_\text{ON}\\
\ket{0_R}_\text{ON}\\
\ket{1_R}_\text{ON}\\
\end{pmatrix} 
= \mathbf C^T \begin{pmatrix}
\ket{0_L}\\
\ket{1_L}\\
\ket{0_R}\\
\ket{1_R}\\
\end{pmatrix}= 
\begin{pmatrix}
\multicolumn{2}{c}{\multirow{2}{*}{\bf A}} & \multicolumn{2}{c}{\multirow{2}{*}{\bf B}}\\
& & & \\
\multicolumn{2}{c}{\multirow{2}{*}{\bf B}} & \multicolumn{2}{c}{\multirow{2}{*}{\bf A}}\\
& & & \\
\end{pmatrix} 
\begin{pmatrix}
\ket{0_L}\\
\ket{1_L}\\
\ket{0_R}\\
\ket{1_R}\\
\end{pmatrix},\label{smeq:ONB}
\end{equation}
where $\mathbf A$ and $\mathbf B$ are general $2\times 2$ matrices. 

Let us consider the state $\ket{i}_\text{ON}$ which is a linear combination of the states $\ket{n} \equiv \ket{a_{L(R)}}$, that can have nonzero overlap $\braket{n| m}$, i.e.,
\begin{equation}
\ket{i}_\text{ON} = \sum \limits_{n}(C^T)_{in} \ket{n} = \sum \limits_{n}C_{ni} \ket{n},\,\, i=1,...,4 \, ,
\label{smeq:ONBi}
\end{equation}
where $C_{ni}$ are coefficients. Next, we impose the  orthonormality condition on these states,
\begin{equation}
\begin{split}
\braket{i|_\text{ON}|j}_\text{ON}& =  \sum \limits_{n, m} C^*_{ni} C_{mj} \braket{n| m}\\
&=  \sum \limits_{n, m} C^*_{ni}\, S_{n m} \, C_{mj} \overset{!}{=} \delta_{ij}\, ,
\end{split}
\label{smeq:ortij}
\end{equation}
where we defined the overlap-matrix $S_{n m} = \braket{n| m}$. Making the choice $C_{nm} = (S^{-1/2})_{nm}$ (used e.g. in Ref.~[\onlinecite{bosco:prb19}]), the transformation matrix is not only Hermitian but also acquires the block-matrix structure of Eq.~\eqref{smeq:ONB} as  will be shown below. 

First the eigenvalue problem of the overlap-matrix will be solved and then the inverse square-root matrix will be calculated via its eigen-decomposition.

In the $L\leftrightarrow R$ symmetric case the aligned overlaps $S_{00}$ and $S_{11}$ are real and the anti-aligned overlaps are related via conjugation $S_{10} = S^*_{01}$. Therefore the overlap-matrix $S_{nm}$ obtains a simple form
\begin{equation}
\mathbf S = \mathbb 1_{4\times 4} + \begin{pmatrix}
0 & 0 & S_{00} & S_{01}\\
0 & 0 & S^*_{01} & S_{11}\\
S_{00} & S_{01} & 0 & 0 \\
S^*_{01} & S_{11} & 0 & 0
\end{pmatrix}\, ,
\end{equation}
where $\mathbb 1_{4\times 4}$ is the $4\times 4$ identity-matrix. We recall that the states $\ket{0_L}$ and $\ket{1_L}$ are orthonormal and the same applies to the right QD. Since the matrix has two identical Hermitian blocks, its eigenvectors are of the following form 
\begin{subequations}
\begin{equation}
{\bf v}^{(1)} = 
\frac{1}{\sqrt{2}} \begin{pmatrix}
v_1\\
v_2\\
v_1\\
v_2
\end{pmatrix}, \hspace{0.4cm}
{\bf v}^{(2)}= 
\frac{1}{\sqrt{2}}\begin{pmatrix}
v_1\\
v_2\\
-v_1\\
-v_2
\end{pmatrix},
\end{equation}
\begin{equation}
{\bf v}^{(3)}= 
\frac{1}{\sqrt{2}} \begin{pmatrix}
-v_2^*\\
v_1\\
-v_2^*\\
v_1
\end{pmatrix},
{\bf v}^{(4)}= 
\frac{1}{\sqrt{2}}\begin{pmatrix}
-v_2^*\\
v_1\\
v_2^*\\
-v_1
\end{pmatrix},
\end{equation}
\label{smeq:cvecs}
\end{subequations}
corresponding to the eigenvalues $1+\lambda_+$, $1-\lambda_+$, $1+\lambda_-$, and $1-\lambda_-$, respectively. The parameters of the eigenvectors and eigenvalues are given by
\begin{subequations}
\begin{equation}
\begin{split}
v_1=\frac{1}{\cal N'}\left(\frac{S_{00}-S_{11}}{2} + |{\rm w}_S|\right),
\hspace{0.5cm}
v_2 = \frac{S_{10}}{\cal N'},\\
\lambda_\pm = \frac{S_{00}+S_{11}}{2} \pm |{\rm w}_S|,
\hspace{1cm}
\end{split}
\label{smeq:ablambda}
\end{equation}
where we used the following definitions:
\begin{equation}
\begin{split}
&{\cal N'}^2 = \left(\frac{S_{00}-S_{11}}{2} + |{\rm w}_S|\right)^2 + |S_{01}|^2,\\
&|{\rm w}_S|^2 =(S_{00}-S_{11})^2/4 + |S_{01}|^2.
\end{split}
\end{equation}
\end{subequations}
The inverse square-root matrix $\mathbf C = \mathbf S^{-1/2}$ is then obtained in the eigendecomposition.

The first block $\mathbf A$ of the transformation matrix $\mathbf C^T$ reads as
\begin{subequations}
\begin{widetext}
\begin{equation}
\begin{split}
\mathbf A &=
\frac 1 2
\begin{pmatrix}
v_1^2 \left(\frac{1}{\sqrt{1-\lambda_+}} + \frac{1}{\sqrt{1+\lambda_+}} \right) + |v_2|^2 \left(\frac{1}{\sqrt{1-\lambda_-}} + \frac{1}{\sqrt{1+\lambda_-}} \right)  & v_1 v_2 \left(\frac{1}{\sqrt{1-\lambda_+}} + \frac{1}{\sqrt{1+\lambda_+}} - \frac{1}{\sqrt{1-\lambda_-}} - \frac{1}{\sqrt{1+\lambda_-}} \right) \\
v_1 v_2^* \left(\frac{1}{\sqrt{1-\lambda_+}} + \frac{1}{\sqrt{1+\lambda_+}} - \frac{1}{\sqrt{1-\lambda_-}} - \frac{1}{\sqrt{1+\lambda_-}} \right)  & v_1^2 \left(\frac{1}{\sqrt{1-\lambda_-}} + \frac{1}{\sqrt{1+\lambda_-}} \right) + |v_2|^2 \left(\frac{1}{\sqrt{1-\lambda_+}} + \frac{1}{\sqrt{1+\lambda_+}} \right) 
\end{pmatrix}\\
&=\begin{pmatrix}
1+ \frac 3 8 \left(v_1^2 \lambda_+^2 + |v_2|^2 \lambda_-^2 \right) & v_1 v_2\, \frac 3 8 \left(\lambda_+^2 - \lambda_-^2 \right) \vspace{0.3cm}\\
v_1 v_2^* \, \frac 3 8 \left(\lambda_+^2 - \lambda_-^2 \right)  &1+ \frac 3 8 \left(v_1^2 \lambda_-^2 + |v_2|^2 \lambda_+^2 \right)
\end{pmatrix} +\mathcal{O} (S_{ab}^3),
\end{split}
\label{smeq:Amat}
\end{equation}
\end{widetext}
where we exploited that $v_1^2 +|v_2|^2 = 1$ and $\lambda_\pm = \mathcal{O} (S_{ab})$ as implied by Eq.~\eqref{smeq:ablambda}. The lowest order corrections to the identity matrix are of second order in $S_{ab}=\braket{a_L|b_R}$. Similarly, one can obtain the matrix $\mathbf B$ and perform the Taylor expansion in $S_{ab}$ as
\begin{widetext}
\begin{equation}
\begin{split}
\mathbf B &=
\frac 1 2
\begin{pmatrix}
v_1^2 \left(-\frac{1}{\sqrt{1-\lambda_+}} + \frac{1}{\sqrt{1+\lambda_+}} \right) + |v_2|^2 \left(-\frac{1}{\sqrt{1-\lambda_-}} + \frac{1}{\sqrt{1+\lambda_-}} \right)  & v_1 v_2 \left(-\frac{1}{\sqrt{1-\lambda_+}} + \frac{1}{\sqrt{1+\lambda_+}} + \frac{1}{\sqrt{1-\lambda_-}} - \frac{1}{\sqrt{1+\lambda_-}} \right) \\
v_1 v_2^* \left(-\frac{1}{\sqrt{1-\lambda_+}} + \frac{1}{\sqrt{1+\lambda_+}} + \frac{1}{\sqrt{1-\lambda_-}} - \frac{1}{\sqrt{1+\lambda_-}} \right)  & v_1^2 \left(-\frac{1}{\sqrt{1-\lambda_-}} + \frac{1}{\sqrt{1+\lambda_-}} \right) + |v_2|^2 \left(-\frac{1}{\sqrt{1-\lambda_+}} + \frac{1}{\sqrt{1+\lambda_+}} \right) 
\end{pmatrix}\\
&=-\frac 1 2 \begin{pmatrix}
v_1^2 \lambda_+ + |v_2|^2 \lambda_- & v_1 v_2 \left(\lambda_+ - \lambda_- \right) \vspace{0.3cm}\\
v_1 v_2^* \left(\lambda_+ - \lambda_- \right)  & v_1^2 \lambda_- + |v_2|^2 \lambda_+
\end{pmatrix} +\mathcal{O} (S_{ab}^3),
\end{split}
\label{smeq:Bmat}
\end{equation}
\end{widetext}
\end{subequations}
which turns out to be of the first order in $S_{ab}$.

Finally, we comment on the well-known case of electrons with separable wave functions, where $S_{01} =0$ and $S_{00} =S_{11} = \mathcal S$, and thus $\lambda_\pm = \mathcal S$.
Substituting this into Eqs.~\eqref{smeq:Amat}~and~\eqref{smeq:Bmat} and using $v_1^2+|v_2|^2=1$, the formula used for the orthonormalized states in Refs.~[\onlinecite{burkard:prb99}, \onlinecite{bosco:prb19}] is recovered.

\section{Single-particle basis for the numerics}
\label{sm:basis}

The eigenvalue problem of Eq.~\eqref{eq:QD} was solved numerically using a finite number of basis states. The eigenstates $\ket a$ of the Hamiltonian in Eq.~\eqref{eq:QD} can be expanded on the product basis of orbital states $\ket{m,n, p}$ and the spin-3/2 eigenstates $\ket j$ as
\begin{equation}
\ket a = \sum \limits_{m,n,p,j} c^{m,n,p}_j \ket{m,n,p,j}\, ,
\end{equation}
where $p$ is the orbital quantum number corresponding to the wire axis, $m$ and $n$ are the orbital quantum numbers corresponding to the transverse direction, $j$ is the eigenvalue of a spin-$3/2$ operator $\hat J_z$, and the expansion coefficients are given by $c^{m,n,p}_j = \braket{m,n,p,j|a}$.

For the analysis of the overlaps in the presence of both electric and magnetic fields we used hard-wall confinement in the transverse directions, i.e.,
\begin{equation}
V_\text{NW} (x,y) = 
\begin{cases}
      0, & \text{if}\ x^2+y^2<R^2 \\
      \infty, & \text{otherwise}
\end{cases}\, ,
\label{eq:VHW}
\end{equation}
with $R$ being the radius. In this case, the basis states can be decomposed into a product form $\ket{m,n,p,j} = \ket{m,n}\ket{p,j}$. For the state corresponding to the transverse directions $\ket{m,n}$ the wave function is given in terms of Bessel-functions of the first kind, 
\begin{equation}
\begin{split}
\braket{r,\phi| m, n} =& \frac{1}{J_{m+1} (x_{m,n})\sqrt{\pi} R}\, \,J_m \left(x_{m,n} \frac{r}{R}\right)\\
&\times
\begin{cases}
      \sqrt 2 \cos (m\phi), & \text{for}\ m>0 \\
      1, & \text{for}\ m=0\\
      \sqrt 2 \sin (m\phi), & \text{for}\  m<0\, ,
\end{cases}
\end{split}
\end{equation}
where $x_{m,n}$ is the n$th$ root of the m$th$ Bessel function $J_m$. The confinement of the QD along the wire ($z$ axis) is assumed to be harmonic, with the corresponding eigenstates given by Hermite polynomials $\text H_p (z)$. The $z$-resolved wave function of the second part $\ket{p,j}$ then becomes
\begin{equation}
\braket{z|p,j} = \frac{\text H_p(z/l_{zj})\exp(-\frac{z^2}{2l_{zj}^2})}{\sqrt{2^p p! l_{zj} \sqrt \pi}} \ket{j}\, ,
\label{eq: hermite}
\end{equation}
where the confinement length is defined with the effective mass $m_{zj}$ as $l^4_{zj} =\hbar^2 a^2/(8 v_B m_{zj})$. The effective mass is obtained by taking the coefficient of the $k_z^2$ term in $\braket{j|H_\text{LK}|j}$, and equating it with $\hbar^2/(2m_{zj})$. Allowing for different confinement lengths $l_{zj}$ in the basis states $\braket{z|p,j}$ for different $j$, we can reduce the off-diagonal elements in the Hamiltonian.
In our calculation for Fig.~\ref{fig:overlaps}, the quantum numbers can take the following values: $m \in \{-2,-1,0,1,2\}$, $n \in \{1,2,3\}$, and $p\in \{0,1,2,3,4\}$.

Calculating the coefficients of the exchange interaction in Fig.~\ref{fig:exchange} required the numerical evaluation of the matrix elements of the Coulomb interaction $\sim 1/|\mathbf r_1 - \mathbf r_2|$ between two-particle basis states. In the basis of Bessel-functions the solution leads to long running times and poor accuracy. However, assuming harmonic confinement in the transversal directions and using the basis of Hermite polynomials in the $x$ and $y$ directions analogous to Eq.~\eqref{eq: hermite} facilitates the analytical calculation of the matrix elements in the transversal directions. For Fig.~\ref{fig:exchange} the quantum numbers of the basis states can take the following value $m \in \{0,1,2\}$, $n \in \{0,1,2\}$ and $p\in \{0,1,2,3,4\}$.

\section{$S_{01}$ as a figure of merit for the effective spin mixing}
\label{sm:S01}

\begin{figure*}
\centering
\includegraphics[width=0.95\textwidth]{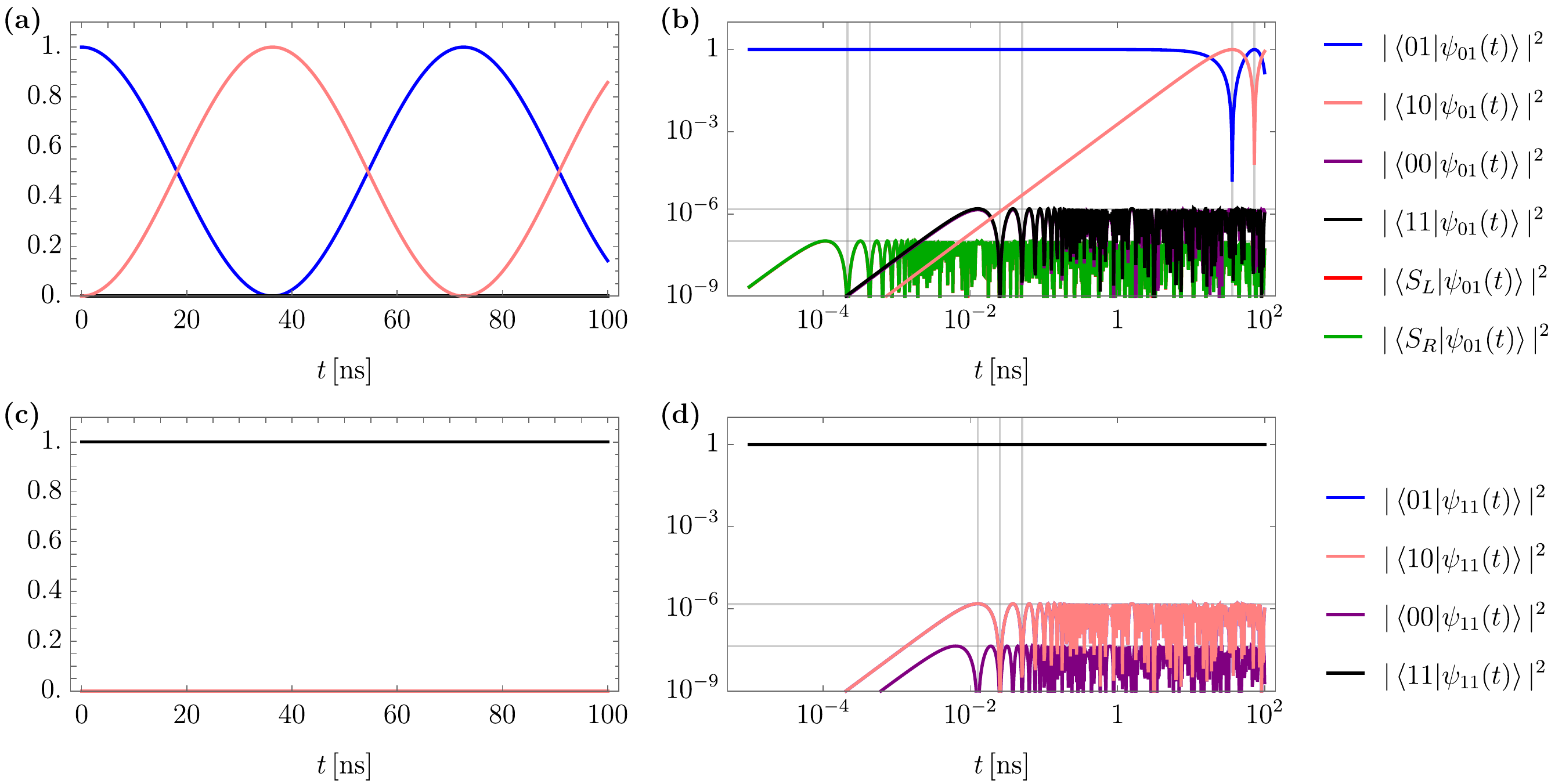}
\caption{Overlaps of the $\ket{\psi_{01}(t)}$ state [(a) and (b)], and the $\ket{\psi_{11}(t)}$ state [(c) and (d)] as a function of time for the low symmetry case, where $\varphi = \pi/8$ and $v_B = 15\,\text{meV}$, for a silicon NW. The time evolution is shown (a) on a liner-linear scale (b) on a log-log scale, with the horizontal lines showing the estimates for the overlaps obtained in Eqs.~\eqref{eq:SWAP2},~\eqref{eq:SWAP3},~and~\eqref{smeq:SWAP2} [(c) and (d) similarly]. Horizontal lines are showing the maximal overlap as a function of time [note that the one corresponding to $|\braket{01|\psi_{11}(t)}|^2 = |\braket{11|\psi_{01}(t)}|^2$ is shown on both (b) and (d)]. Overlaps with different basis states are oscillating with the half cycle duration of $\sim h/U$ for $|\braket{S_L|\psi_{01}(t)}|^2$, $\sim h/\Delta_z$ for $|\braket{11|\psi_{01}(t)}|^2$, and $\sim h/(2\Delta_z)$ for $|\braket{00|\psi_{11}(t)}|^2$ as illustrated by the vertical lines in (b) and (d).}
\label{smfig:SWAPerrors}
\end{figure*}

In the main text we argued that the quantity $S_{01}$ is a good measure for the (unwanted) mixing of effective spins (qubits) in the absence of a twofold symmetry. Here we will show that $S_{01}$ can be expressed in terms of the anisotropy terms $\Delta_{x,y}$ of the DQD Hamiltonian as well as the off-diagonal exchange matrix elements such as $J_{xz}$.

To this end we focus on our particular example of a silicon NW with $[001]$ growth direction and in the presence of a perpendicular magnetic field and perform 
the symmetry decomposition given in Eq.~\eqref{eq:decomp} of the corresponding single QD Hamiltonian in Eq.~\eqref{eq:QD} according to the mirror symmetry $S_{2B}$. The symmetry breaking part $H_\text{SO}$ contains the terms shown in Eqs.~\eqref{eq:anissoia}-\eqref{eq:anissoic}. From this decomposition we derive an effective $2\times 2$ Hamiltonian describing the lowest energy (or qubit) subspace. A convenient way to do this is to find the eigenstates of the high-symmetry part $H_0$ and perform an exact Schrieffer-Wolff transformation. This leads to
\begin{equation}
H^{2\times 2}_\text{QD} =H^{2\times 2}_{0}+H^{2\times 2}_\text{SO}= \frac{\Delta'_0}{2} \sigma'_z + \text{Re}(\nu) \sigma'_x + \text{Im}(\nu) \sigma'_y
\label{smeq:H2x2}
\end{equation}
for the lowest $2\times 2$ block of $H_\text{QD}$, which is decoupled from the rest of the states. The Pauli matrices above are defined as $\sigma'_z = \ket{0+}\bra{0+} - \ket{1-}\bra{1-}$, with the states $\ket{0+}$ and $\ket{1-}$ being eigenstates of the symmetry operator $D(S_{2B})$. The energy splitting between the states $\ket{0+}$ and $\ket{1-}$ is $\Delta_0'$ and the symmetry breaking part of the Hamiltonian $H^{2\times 2}_\text{SO}$ is proportional to $\nu = \braket{{1-}|H^{2\times 2}_\text{SO}|0+}=\braket{{1-}|H^{2\times 2}_\text{QD}|0+}$. Furthermore, $\text{Re}(\nu)$ and $\text{Im}(\nu)$ correspond to the real and imaginary parts of $\nu$, respectively.

Even though the coupling $\nu$ cannot be expressed in a simple form generally, for the NW system considered in Sec.~\ref{sec:NWexchange} one can simply extract relations for $\nu$ in two special cases, without performing an explicit Schrieffer-Wolff transformation. These are:

(i) If $E_B = 0$, the symmetry breaking part contains only Eqs.~\eqref{eq:anissoia}~and~\eqref{eq:anissoib} and therefore $H_\text{SO} \sim \sin{(4\varphi)}$. Consequently, the coupling associated with the cubic anisotropy is $\nu \sim \sin{(4\varphi)}$. The values of $\varphi$ where the coupling vanishes correspond to high-symmetry directions in the system where $S_{2B}$ is a symmetry of the Hamiltonian $H_\text{QD}$.

(ii) If we treat the bulk Hamiltonian $H_\text{LK}+H_Z$ in the axial approximation (i.e., $\Delta \gamma = q = 0$), the coupling induced by the electric field is $\nu \sim e E_B$, similarly to the case of Rashba SOI. Therefore, we associate this effect to  DRSOI\cite{kloeffel:prb11,kloeffel:prb18}.

Diagonalizing the effective $2\times 2$ Hamiltonian of Eq.~\eqref{smeq:H2x2}, we recover  the eigenstates of $H_\text{QD}$ as
\begin{subequations}
\begin{equation}
\ket{0} = \frac{\Delta_0+\Delta_0'}{2 {\cal N}} \ket{0+} + \frac \nu {\cal N} \ket{1-}\, ,
\end{equation}
\begin{equation}
\hspace{1.1cm} \ket{1} =  - \frac{\nu^*} {\cal N} \ket{0+} +\frac{\Delta_0+\Delta_0'}{2{\cal N}} \ket{1-}\, ,
\end{equation}
\label{smeq:eig01}
\end{subequations}
corresponding to the energies $+\Delta_0/2 = \sqrt{{\Delta'}_0^2/4+|\nu|^2}$ and $-\Delta_0/2$, respectively. The normalization factor is ${\cal N} = \sqrt{(\Delta_0+\Delta_0')^2 /4 + |\nu|^2}$.

Moving on to the DQD problem, we introduce the low-energy basis $\ket{0_{L(R)}+}$ and $\ket{1_{L(R)}-}$ associated to the left (right) QDs and define the overlaps
\begin{subequations}
\begin{equation}
s_0=\braket{{0_L+}|0_R+}
\end{equation}
\begin{equation}
s_1= \braket{{1_L-}|1_R-},
\end{equation}
\label{smeq:olapspm}
\end{subequations} 
which are in general nonzero, whereas the anti-aligned overlaps vanish, i.e., $\braket{{1_L-}|0_R+}=\braket{{0_L+}|1_R-}=0$ due to the symmetry properties of the basis states. The system we consider in Sec.~\ref{sec:NWexchange} is $L\leftrightarrow R$ symmetric, implying that the quantities $s_0$ and $s_1$ are real. Exploiting the relations between the basis states Eq.~\eqref{smeq:olapspm} , we write the overlap between eigenstates $\ket{0_L}$ and $\ket{1_R}$ as
\begin{equation}
S_{01} = \braket{0_L|1_R} = \frac{\nu^*}{{\cal N}}  \frac{\Delta_0+\Delta_0'}{2{\cal N}}  \, (s_1-s_0)\, .
\end{equation}
Due to the cylindrical symmetry of the NW the confinement respects the symmetry $S_{2B}$, if the magnetic field is applied perpendicularly to the wire. The induced Zeeman splittings of the left QD reads
\begin{eqnarray}
\begin{split}
\frac{\Delta^L_x-i \Delta^L_y}{2} &= \frac{\nu^*}{{\cal N}}  \frac{\Delta_0+\Delta_0'}{2{\cal N}} \\
&\times \left[  \braket{{1_L-}|\delta V_L|1_L-}-\braket{{0_L+}|\delta V_L|0_L+}\right]
 \, ,
\end{split}
\end{eqnarray}
where $\delta V_L = V_\text{DQD} -V_L$, and the corrections due to the orthogonalization of the left and right bases are neglected. At last, we show the formula for the coupling matrix element $\braket{T_+|C|T_0}$ (to lowest order in the overlaps)
\begin{eqnarray}
\begin{split}
 \frac{J_{xz}^s - iJ_{yz}^s}{2\sqrt 2}= \frac{\nu^*}{{\cal N}} &\left( \frac{\Delta_0+\Delta_0'}{2{\cal N}} \right)^3 \sqrt 2 \\
\times &\big[\braket{{0_L+}, {1_R-}|C|0_L+, 1_R-} \\
&- \braket{{0_L+}, {0_R+}|C|0_L+, 0 _R+}\big]
 \, .
\end{split}
\end{eqnarray}
Importantly, the overlap $S_{01}$ and anisotropic couplings share the prefactor $\nu^* /{\cal N}$, which is typically a small parameter. Therefore, the simple quantity $S_{01}$ does not only show the symmetry properties  of $H_\text{DQD}$ but can also be used to study the competition of the two main spin mixing effects, the cubic anisotropy and DRSOI.

\section{Slow $\sqrt{\text{SWAP}}$ gates and anisotropy-limited fidelity}
\label{sm:timevo}

\begin{figure*}
\centering
\includegraphics[width=\textwidth]{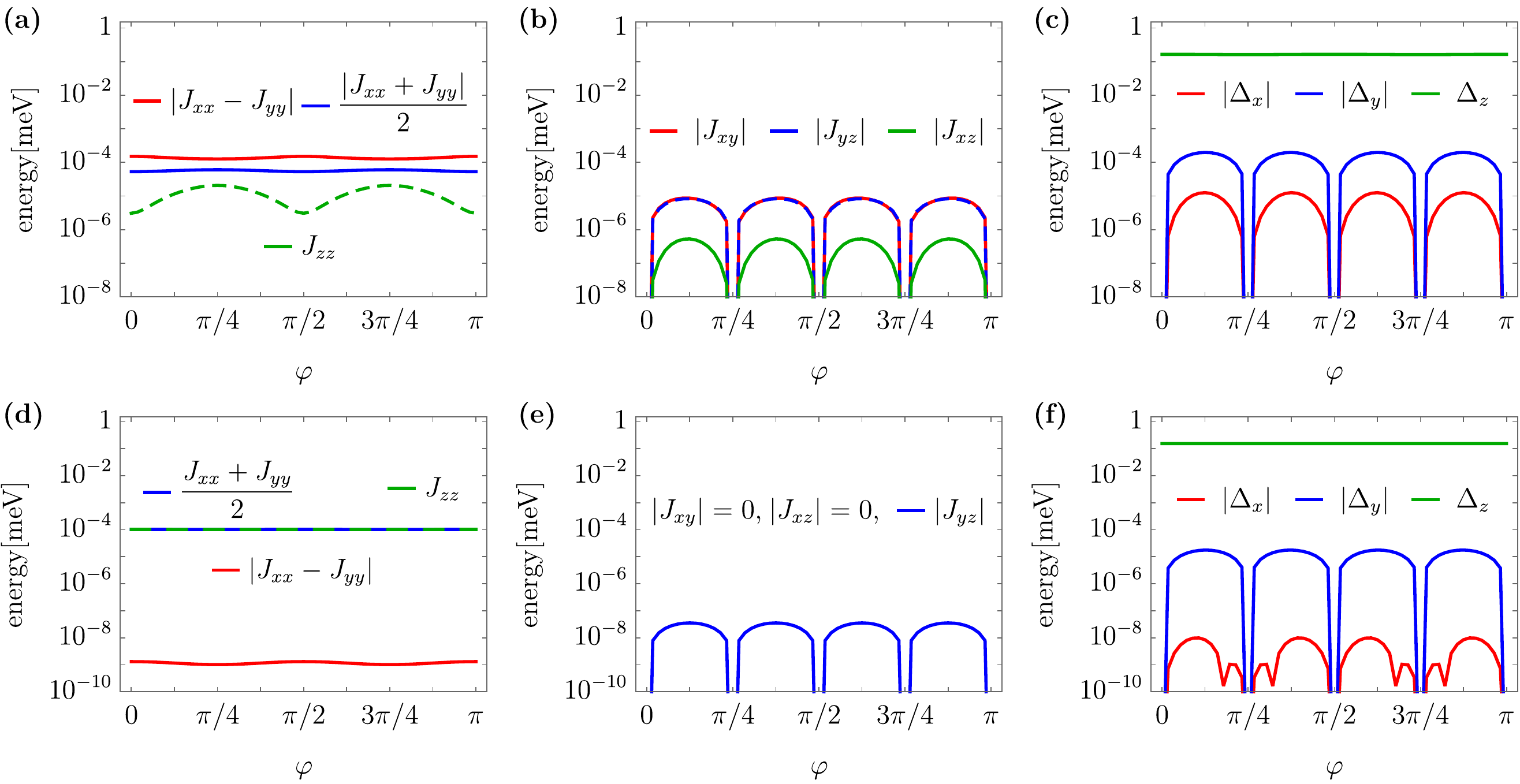}
\caption{(a)-(c) Coefficients characterizing the exchange interaction and the single particle Hamiltonian as a function of the magnetic field direction $\varphi$ for a silicon NW. For the numerical simulation the following parameters were used: double dot distance $2a = 30\, \text{nm}$; barrier height $v_\text B = 15\, \text{meV}$; magnetic field $B = 1\, \text T$; harmonic potential in the transverse directions with a confinement length of $2l_T = 8\, \text{nm}$. (d)-(f) Coefficients characterizing the exchange interaction as a function of the magnetic field direction $\varphi$ for a Ge/Si core/shell NW. For the numerical simulation the following parameters were used: double dot distance $2a = 60\, \text{nm}$; barrier height $v_\text B = 15\, \text{meV}$; magnetic field $B = 0.5\, \text T$; relative shell thickness\cite{kloeffel:prb14} $(R_s - R_c)/R_c = 0.2$; harmonic potential in the transverse directions with a confinement length of $2l_T = 8\, \text{nm}$.}
\label{smfig:exchangesm}
\end{figure*}

In Sec.~\ref{subsec:SWAP} we considered the time evolution of the states $\ket{\psi_{01} (t)}$ and $\ket{\psi_{11} (t)}$, identified the couplings leading to the largest undesired overlaps which provided good estimates for the error rates $1-\mathcal F_{01}$ and $1-\mathcal F_{11}$ for the $\sqrt{\text{SWAP}}$ gate. We found that the error rate $1-\mathcal F_{01}$ is orders of magnitudes higher than the one corresponding to the $\ket{11}$ state, due to the possibility of tunneling to a doubly occupied state.

In this appendix we discuss a parameter regime for the $\sqrt{\text{SWAP}}$ gate where the two fidelities are limited by the same transition probability that is set by the anisotropic couplings. This is the regime where the potential barrier is high enough, e.g., $v_B=15\,\text{meV}$ in Fig.~\ref{smfig:SWAPerrors}, while the rest of the parameters were set to be  identical to the case of Figs.~\ref{fig:exchange}~and~\ref{fig:SWAPerrors}. The corresponding error rate for the $\sqrt{\text{SWAP}}$ gate is obtained as
\begin{equation}
1-\mathcal F \sim 2 |\braket{01|\psi_{11}(\tau_s)}|^2 \sim 2 \frac{\Delta_{x}^2+\Delta_{y}^2}{\Delta_z^2} \sim 3\cdot 10^{-6}\, ,
\label{smeq:SWAP}
\end{equation}
where $\mathcal F = \mathcal F_{01} = \mathcal F_{11}$ is the gate fidelity that is independent from the input state. When the potential barrier is increased, the singlet triplet splitting is reduced, increasing the $\sqrt{\text{SWAP}}$ operation time by two orders of magnitude to $\tau_s \sim 18\, \text{ns}$.

For large enough potential barrier $v_B$, Eq.~\eqref{eq:SWAP1} loses its validity since in general the transition probability is set by the tunnel coupling $\braket{S_L|H_\text{DQD}^{6\times 6}|S}$, i.e.,
\begin{equation}
|\braket{S_L|\psi_{01}(\tau_s)}|^2 \sim 2\frac{|\braket{S_L|H_\text{DQD}^{6\times 6}|S}|^2}{U^2}\sim 10^{-7}\, .
\label{smeq:SWAP2}
\end{equation}
This is an order of magnitude smaller than the leading correction to the error rate. Therefore, we conclude that the anisotropic corrections are influencing the $\sqrt{\text{SWAP}}$ gate fidelities only for very slow gates.

\section{Comparison of the exchange interaction between silicon and Ge/Si core/shell NWs}
\label{sm:SiGe}

In this appendix we present the role of the anisotropy parameters $\Delta \gamma/\overline \gamma$ and $q/|\kappa|$ for the exchange interaction and, in particular, compare silicon with Ge/Si NWs in the absence of electric fields, i.e., $\mathbf E = 0$. In Fig.~\ref{fig:exchange} we saw for the case of silicon that the cubic anisotropy renders the exchange interaction anisotropic and introduces off-diagonal terms $J_{xz},J_{yz}$, if the magnetic field is applied in a low-symmetry direction. As we have seen in App.~\ref{sm:S01}, these anisotropic effects disappear in the axial approximation since the symmetry breaking parts of the Hamiltonian, Eqs.~\eqref{eq:anissoia}~and~\eqref{eq:anissoib} are proportional to $\Delta \gamma$ and $q$, respectively.

Next, given the recent experimental interest, we consider a Ge/Si core/shell NW, with $[001]$ growth direction. The coordinate axes were chosen identically to the case of silicon ($x,y,z$ correspond to the $[100]$, $[0 1 0]$ and $[001]$ crystallographic axes, respectively) in which case the Bir-Pikus Hamiltonian becomes
\begin{equation}
H^{[001]}_\text{BP} = b (\epsilon_{zz}-\epsilon_\perp) J_{z}^2\, ,
\label{smeq:BP}
\end{equation}
where we omitted a constant part, $b = -2.5\, \text{eV}$, while the values of the strain ($\epsilon_{zz}$ and $\epsilon_\perp$) as a function of relative shell thickness $(R_s - R_c)/R_c = 0.2$ are taken from Ref.~[\onlinecite{kloeffel:prb14}].

We performed the calculation similarly to Sec.~\ref{subsec:anis}. For the case of a silicon NW in Figs.~\ref{smfig:exchangesm}(a)-(c) we used the same parameters as in Sec.~\ref{subsec:anis}, the only difference being that we used a significantly higher potential barrier, $v_B = 15\, \text{meV}$. For the case of the Ge/Si core/shell NW [see Figs.~\ref{smfig:exchangesm}(d)-(f)], the shell is taken into account via the strain term in the Hamiltonian of Eq.~\eqref{smeq:BP}, and the Luttinger parameters are $\gamma_1 = 13.38$, $\gamma_2 = 4.24$, $\gamma_3 = 5.69$, $\kappa=3.41$, and $q = 0.06$\cite{winkler:book}. Note that while the anisotropy parameters for silicon are $\Delta \gamma / \bar \gamma = 1.1$ and $q/|\kappa| = 0.024$, for germanium the same quantities are significantly smaller, i.e., $\Delta \gamma / \bar \gamma = 0.28$ and $q/|\kappa| = 0.018$.

A remarkable reduction can be observed in the anisotropic exchange terms for a Ge/Si core/shell NW (see Fig.~\ref{smfig:exchangesm}), compared to the case of silicon NWs. The parameters are set such that the Zeeman splitting and the diagonal exchange matrix elements are of the same order for the two materials. One can directly see that although in the case of silicon NW the off-diagonal terms can be comparable to the diagonal ones, they almost disappear (at least they are below the accuracy of our numerics) for the strained Ge/Si core/shell NW. 


\newpage 

\bibliographystyle{unsrt}

\begin{thebibliography}{}
\bibitem{footnote:cubcorr}{$K(\mathbf k, \hat{\mathbf J})= \frac{\hbar^2}{2m}\left[-\frac{3}{2}k^2+ \frac{6}{5}(k_{x'}^2 \hat J_{x'}^2+k_{y'}^2 \hat J_{y'}^2+k_{z'}^2 \hat J_{z'}^2)\right.$ $\left.- \frac{8}{5}( \{k_{x'},k_{y'}\}\{\hat J_{x'},\hat J_{y'}\} +\{k_{y'},k_{z'}\}\{\hat J_{y'},\hat J_{z'}\}\right.$ $\left. + \{k_{z'},k_{x'}\}\{\hat J_{z'},\hat J_{x'}\}) \right]$, where the anti-commutator is defined as $\{A,B\} = \frac{1}{2}(AB+BA)$.}
\bibitem{footnote:LKtypo}{Note that the corresponding equations (64) and (65) of Ref.~[\onlinecite{luttinger:pr56}] contain a typo in the first terms of the Hamiltonians, i.e. $k^2\rightarrow k^2/2$. Furthermore, in equation (65) of Ref.~[\onlinecite{luttinger:pr56}] in the Zeeman term $(\kappa-\overline \gamma/2)$ should be repalced by $(\kappa+\overline \gamma/2)$. These corrections are then consistent with the equations (45), (69), and the rest of Ref.~[\onlinecite{luttinger:pr56}].}
\bibitem{footnote:aniszeeman}{The anisotropic Zeeman term reads as $2 q \mu_\text B \mathbf B \cdot \mathbf{\hat{\mathcal J}} = 2 q \mu_\text B \left(B_{x'} \hat J_{x'}^3+B_{y'} \hat J_{y'}^3+B_{z'} \hat J_{z'}^3 \right)$ where the axes $x',y',z'$ are along the crystallographic axes $\braket{100}$.}
\bibitem{footnote:modulo}{Within the framework of this paper, we define the modulo function with shifted range, i.e., $\text{mod}_N(x) \in [-N/2,\,N/2)$, for any $x\in {\mathbb R}$. This is related to the conventional definition via $\text{mod}_N (x)=\text{mod}'_N (x+N/2)-N/2$, where $\text{mod}'_N (x) \in [0,N)$.}
\bibitem{footnote:partexchsym}{In our calculations we consider homogeneous electric and magnetic fields. Therefore $H_{QD} - V_{QD}$ [see Eq.~\eqref{eq:QD}] is identical for both QDs and Eq.~\eqref{eq:DQD} invariant under particle exchange.}
\bibitem{footnote:gauge}{Note that the matrix elements of the Hamiltonian in Eq.~\eqref{eq:H11long} are not independent of the relative phase of the basis states, and therefore only the absolute values of the off-diagonal exchange couplings and induced Zeeman splittings are shown in Figs.~\ref{fig:exchange}~and~\ref{smfig:exchangesm}.}
\bibitem{footnote:growthdirection}{We refer here to the direction along the NW as the growth direction.}
\end{thebibliography}

\end{document}